\documentclass[12pt,letterpaper]{article}

\usepackage{verbatim}
\usepackage{float}
\usepackage{booktabs}
\usepackage{amsmath}
\usepackage{amssymb}
\usepackage{graphicx}
\usepackage{setspace}
\usepackage{natbib}
\usepackage{algorithm}
\usepackage{algorithmic}
\usepackage{xcolor, colortbl,epstopdf}

\newtheorem{example}{Example}

\definecolor{lightgray}{gray}{0.9}

\bibpunct{(}{)}{;}{a}{,}{,}

\newcommand{\ba}{\mathbf{a}}
\newcommand{\bA}{\mathbf{A}}

\newcommand{\bB}{\mathbf{B}}
\newcommand{\bc}{\mathbf{c}}
\newcommand{\bC}{\mathbf{C}}

\newcommand{\bff}{\mathbf{f}}

\newcommand{\bG}{\mathbf{G}}
\newcommand{\bH}{\mathbf{H}}
\newcommand{\bh}{\mathbf{h}}

\newcommand{\bK}{\mathbf{K}}

\newcommand{\bM}{\mathbf{M}}
\newcommand{\bO}{\mathbf{O}}

\newcommand{\bQ}{\mathbf{Q}}

\newcommand{\bs}{\mathbf{s}}
\newcommand{\bS}{\mathbf{S}}
\newcommand{\bu}{\mathbf{u}}
\newcommand{\bU}{\mathbf{U}}
\newcommand{\bv}{\mathbf{v}}

\newcommand{\bx}{\mathbf{x}}
\newcommand{\bX}{\mathbf{X}}
\newcommand{\by}{\mathbf{y}}

\newcommand{\bV}{\mathbf{V}}
\newcommand{\bW}{\mathbf{W}}
\newcommand{\bz}{\mathbf{z}}

\newcommand{\vect}[1]{\boldsymbol #1}

\newcommand{\valpha}{\vect{\alpha}}
\newcommand{\vbeta}{\vect{\beta}}

\newcommand{\vgamma}{\vect{\gamma}}

\newcommand{\vLambda}{\vect{\Lambda}}
\newcommand{\vmu}{\vect{\mu}}
\newcommand{\vphi}{\vect{\phi}}
\newcommand{\vPhi}{\vect{\Phi}}

\newcommand{\vepsilon}{\vect{\epsilon}}

\newcommand{\vOmega}{\vect{\Omega}}
\newcommand{\vSigma}{\vect{\Sigma}}
\newcommand{\vsigma}{\vect{\sigma}}
\newcommand{\vPsi}{\vect{\Psi}}
\newcommand{\vpsi}{\vect{\psi}}

\newcommand{\vtheta}{\vect{\theta}}

\newcommand{\e}{\text{e}}
\newcommand{\gvn}{\,|\,}
\renewcommand{\epsilon}{\varepsilon}
\renewcommand{\hat}{\widehat}

\renewcommand{\leq}{\leqslant}

\newcommand{\distn}[1]{\mathcal{#1}}
\newcommand{\matlab}{\mathrm{M}\mathrm{{\scriptstyle ATLAB}}}

\begin{document}



\title{High-Dimensional Conditionally Gaussian State Space Models with Missing Data\thanks{A previous version of the paper was circulated under the title ``Efficient Estimation of State-Space Mixed-Frequency VARs: A Precision-Based Approach". We thank Otilia Boldea, Francis Diebold, Gary Koop, James Mitchell and Michele Piffer for many constructive comments and valuable suggestions. 
}}
\author{Joshua C. C. Chan \\
Purdue University \and Aubrey Poon \\
 \"Orebro University \and Dan Zhu \\
 Monash University}

\date{January 2023}

\maketitle

\begin{abstract}

\noindent We develop an efficient sampling approach for handling complex missing data patterns and a large number of missing observations in conditionally Gaussian state space models. Two important examples are dynamic factor models with unbalanced datasets and large Bayesian VARs with variables in multiple frequencies. A key insight underlying the proposed approach is that the joint distribution of the missing data conditional on the observed data is Gaussian. Moreover, the inverse covariance or precision matrix of this conditional distribution is sparse, and this special structure can be exploited to substantially speed up computations. We illustrate the methodology using two empirical applications. The first application combines quarterly, monthly and weekly data using a large Bayesian VAR to produce weekly GDP estimates. In the second application, we extract latent factors from unbalanced datasets involving over a hundred monthly variables via a dynamic factor model with stochastic volatility.



\bigskip


\bigskip


\end{abstract}

\thispagestyle{empty}

\newpage

\section{Introduction}
Large-scale time-series models are increasingly used in empirical macroeconomics to exploit the wide availability of large datasets. This trend promises a more timely and comprehensive analysis but also brings new challenges. First, large datasets are typically compiled from multiple sources, and consequently, they often involve complex missing data patterns. One prominent example is mixed-frequency data to incorporate real-time information, as opposed to the traditional approach of using only variables at the same (lower) frequency. For instance, \citet{ADS09} combine daily, weekly, monthly and quarterly variables to construct a business conditions index to track economic activity. \citet{SS15} use GDP, which is only available quarterly, and other quarterly and monthly variables to obtain GDP estimates at the monthly frequency. In both cases, the high-frequency observations of the low-frequency variables are treated as missing data. As such, there are a large number of missing observations.

Second, extracting information from large datasets generally requires large-scale time-series models. Factor models have been the workhorse for this purpose, and thanks to the seminal work of \citet*{BGR10} and \citet{koop13}, large Bayesian vector autoregressions (VARs) have now become a popular alternative. In addition, since there is a large body of empirical evidence that shows allowing for various flexible features, such as heteroskedasticity, heavy-tailed distributions and outliers detection, are vitally important for improving in-sample model-fit, and out-of-sample forecast performance \citep[see, e.g.,][]{clark11, DGG13, CP16, SW16, chan20}, these features are increasingly incorporated into dynamic factor models and large Bayesian VARs \citep[recent examples include][]{KH20, ADP21, CCMM22}. While there are many recent advances in speeding up the estimation of these flexible large-scale models with complete data, efficient algorithms that can handle complex missing data patterns with a large number of missing observations are by comparison underdeveloped.


We tackle these challenges by developing an efficient sampling approach for drawing all the missing observations in one step. To make our approach widely applicable, it is developed under a general framework of conditionally Gaussian state space models. As such, it applies to many of the popular large-scale models, such as dynamic factor models with stochastic volatility or mixed-frequency VARs with non-Gaussian errors. In addition, the setup can easily handle a wide variety of complex missing data patterns, including unbalanced panels, mixed-frequency settings, and a `ragged edge' at the end of the sample due to non-synchronous data releases. Thanks to the modular nature of Markov chain Monte Carlo (MCMC) methods, the proposed approach can be straightforwardly implemented in conjunction with any efficient samplers for conditionally Gaussian state space models with complete data. Our paper therefore complements existing works on fast estimation of flexible large-scale models and extend them to missing data settings.

The key insight underlying the proposed approach is that the joint distribution of all the missing observations conditional on the observed data (and other model parameters and latent variables) is Gaussian. Furthermore, the precision matrix (i.e., inverse covariance matrix) of this conditional distribution is sparse---in fact, for many of the common missing data patterns, it is banded, i.e., it is sparse, and its non-zero elements are arranged along a diagonal band. These special structures can be exploited to vastly speed up computations. In particular, the precision-based sampler of \citet{CJ09} can be applied to draw all the missing observations in one step. This approach is much more efficient compared to standard Kalman-filter-based methods, especially when there are a large number of missing observations or when the state vector is high-dimensional.\footnote{The precision-based sampling approach of \citet{CJ09} and \citet{MMP11} are designed for linear Gaussian state space models with complete data. It builds upon earlier work on Gaussian Markov random fields \citep{rue01} and nonparametric regression \citep{CJ06,CGJ09}. Due to its ease of implementation and computational efficiency, this approach is increasingly used in a wide range of empirical applications. Recent examples include modeling trend inflation \citep{CKP13,chan17,Hou20}, time-varying Phillips curve \citep{Fu20} and dividend growth \citep{PST23}; estimating the output gap \citep{GC17b,GC17}; macroeconomic forecasting \citep{CP16, CHP20}; and fitting various moving average models \citep{chan13, CEK16, DK20, ZCC20} and dynamic factor models \citep{KS19, BK21}.} In addition, it is straightforward to implement: one only needs to partition the data vector into observed and missing data by defining some appropriate selection matrices. The proposed approach can easily handle many complex missing data patterns, such as settings with variables in multiple frequencies.

In addition, our setup can also accommodate settings in which additional information is available to help sharpen inference on the missing observations. This feature is crucial in mixed-frequency applications where linear combinations of high-frequency missing observations need to be mapped to the observed values of low-frequency variables. Our paper is related to the recent work by \citet{EKMN20} and \citet{HS21}, who also consider a precision-based sampling approach for settings with missing observations. However, they focus on dynamic factor models and the latter does not consider mixed-frequency settings. In contrast, our approach is more general and is applicable to any conditionally Gaussian state space models under a wide variety of missing data patterns.

We conduct a series of Monte Carlo experiments to illustrate the numerical accuracy and computation speed of the proposed precision-based approach. In particular, we estimate mixed-frequency VARs using the proposed samplers and standard filtering methods under a variety of settings. We show that the proposed precision-based approach is much more computationally efficient compared to traditional Kalman-filter-based methods. In addition, it scales well to high-dimensional settings, making it possible to estimate VARs with a large number of low-frequency variables. 

To demonstrate the versatility of the proposed precision-based approach, we consider two empirical macroeconomic applications with widely different missing data patterns. In the first application, we use a large mixed-frequency Bayesian VAR with stochastic volatility to generate weekly estimates of real GDP. These high-frequency GDP estimates are useful for a range of purposes, such as monitoring the current state of the economy and delivering timely nowcasts of key macroeconomic variables. To obtain the weekly GDP estimates, we fit a Bayesian VAR using 22 variables in 3 different frequencies: 16 weekly variables, 5 monthly variables and the quarterly real GDP. All variables are modeled at the weekly frequency, and the weekly observations of the monthly and quarterly variables are treated as missing data. Even though the missing data pattern is complex---e.g., there are different numbers of weeks in different months and quarters---and there are a large number of missing observations, the proposed approach is computationally efficient and easy to implement. 

In the second application, we use a dynamic factor model with stochastic volatility to extract latent factors in real-time from the FRED-MD datasets of \citet{MN16}. Each data vintage of FRED-MD contains 128 monthly variables, but many have missing values from two sources: missing observations at the beginning of the sample for some recently constructed variables and missing values at the end of the sample due to publication lags. We implement the proposed approach to sample the missing observations under the dynamic factor model and obtain the latent factors. Our results show that the first factor tracks the broad economic conditions well, even during the pronounced downturn at the onset of the COVID-19 pandemic and the subsequent rebound. In addition, our results suggest that using only variables without missing values can potentially misrepresent the dynamics of the latent factors, highlighting the importance of incorporating the information from variables with missing values.

The remainder of the paper is organized as follows. Section \ref{s:methodology} discusses the proposed precision-based sampling approach for drawing the missing observations in a general state space framework. Section~\ref{s:MC} conducts a series of Monte Carlo experiments comparing the proposed sampling approach against standard Kalman-filter based techniques in a variety of mixed-frequency settings. Section~\ref{s:applications} demonstrates how the proposed sampling approach can be applied to two popular empirical macroeconomic applications. Finally, Section~\ref{s:conclusion} concludes.

\section{A General State Space Framework} \label{s:methodology}

This section introduces the proposed precision-based approach for sampling the missing data conditional on a variety of information sets under a general state space framework. More specifically, we first derive the joint conditional distribution of the missing data given the observed data and other model parameters, which we show is Gaussian. We then discuss an efficient algorithm to generate samples from this typically high-dimensional Gaussian distribution. In addition, since in many applications, such as mixed-frequency settings, one has additional information on the missing data, we demonstrate how this additional information can be incorporated to update the conditional distribution of the missing data.

\subsection{The Conditional Distribution of the Missing Data}

Our general setup is the following conditionally Gaussian state space model for an $n\times 1$ vector of variables $\by_t=(y_{1,t},\ldots, y_{n,t})'$ over $t=1,\ldots, T$:
\begin{align}
	\by_t & = \bW_t\valpha_t + \bX_t\vbeta + \vepsilon_t, &\vepsilon_t&\sim\distn{N}(\mathbf{0}_n,\vSigma_t), \label{eq:obs} \\
	\valpha_t & = \vgamma + \vPhi_1\valpha_{t-1} + \cdots + \vPhi_q\valpha_{t-q} + \vepsilon_t^{\valpha}, 
	&	\vepsilon_t^{\valpha}& \sim \distn{N}(\mathbf{0}_k,\vOmega_t), \label{eq:state}
\end{align}
where $\mathbf{0}_m$ denotes an $m\times 1$ vector of zeros, $\vbeta $ is a vector of time-invariant parameters, $\valpha_t$ is a vector of time-varying parameters, $\vSigma_t$ and $\vOmega_t$ are the covariance matrices for the observation and state equations, respectively. The covariate matrices $\bW_t$ and $\bX_t$ could include lagged values of $\by_t$. This framework encompasses a wide range of commonly-used models, including dynamic factor models and vector autoregressions. 

Note that it also includes many different types of error processes as special cases. For instance, one can specify $\vSigma_t$ as the multivariate stochastic volatility processes in \citet{CS05}, \citet{Primiceri05}, \citet{CCM16} or \citet{kastner19}. In addition, one can also consider various types of non-Gaussian errors, such as the $t$ distribution by setting $\vSigma_t = \lambda_t\bQ$, where $\bQ$ is a covariance matrix and $\lambda_t\sim\distn{IG}(\nu/2, \nu/2)$, or an outlier component of the type in \citet{SW16} by specifying $\vSigma_t = o_t^2\bQ$, where $o_t$ follows a 2-part distribution with a point mass at 1 and a uniform distribution on the interval $(2,10)$. Naturally, any combination of the above multivariate stochastic volatility processes or non-Gaussian errors, such as those in \citet{chan20} and \citet{CCMM22}, is also possible.

We are interested in settings in which some elements of $\by_t$ are missing. More specifically, partition $\by_t$ into two subvectors, $\by_t^{o}$ and $\by_t^{m}$, where $\by_t^o$ is an $n_t^o$-vector of observed variables and $\by_t^m$ is an $n_t^m$-vector of missing variables such that $n_t^o+n_t^m=n$. Note that here $n_t^o$ and $n_t^m$ can be time-varying, and hence this setup can accommodate a wide range of missing data patterns, such as unbalanced panels and ragged edge. In addition, for settings with variables of mixed frequencies, it is common to express the time index in the highest frequency and treat some of the low-frequency variables as missing. For example, in models with both monthly and quarterly variables, the monthly values of the quarterly stock variables are only observed every 3 months and the rest are treated as missing.\footnote{For flow variables, their observed values can be viewed as additional information that can be mapped to the missing high-frequency values; this case will be further discussed in the next subsection.} Finally, let $N^o = \sum_{t=1}^Tn_t^o $ and $N^m = \sum_{t=1}^Tn_t^m $ denote the total numbers of observed and missing values with $N^o + N^m = Tn$. For later reference, stack $\by = (\by_1',\ldots, \by_T')'\in \mathbb{R}^{Tn}$, $\by^o = (\by_1^{o\prime},\ldots,\by_T^{o\prime})' \in\mathbb{R}^{N^o}$ and $\by^m = (\by_1^{m \prime},\ldots,\by_T^{m \prime})'\in\mathbb{R}^{N^m}$  vectors. 

One popular approach to handle the missing observations $\by^m$ is to treat them as latent variables to be augmented or sampled. This is typically done using standard Kalman filtering and smoothing algorithms. However, the main drawback of this approach is that it tends to be computationally intensive in high-dimensional settings when there are a large number of missing observations. This significant computational burden is a key obstacle in practice for using high-dimensional state space models with missing data, despite the increasing popularity of large-scale dynamic factor models and VARs. In addition, when the missing data pattern is complex, the implementation of Kalman filter based algorithms also becomes more cumbersome. To overcome these computational and implementation issues, we develop an efficient method to jointly sample $\by^m$ given $\by^o$ and other model parameters and latent variables, which we denote as $\vtheta$. The proposed method is conceptually simply and easy to implement, even with complex missing data patterns. 

In what follows, we first derive the joint conditional distribution of $\by^m$. To that end, we write $\by$ in terms of $\by^o$ and $\by^m$:
\begin{equation} \label{eq:selection}
	\by = \bS^o\by^{o} + \bS^{m}\by^{m},
\end{equation}
where $\bS^o$ and $\bS^m$ are, respectively, $Tn \times N^o$ and $Tn \times N^m$ selection matrices. In particular, each column of $\bS^o$ and $\bS^m$ contains only one element that is 1, and all other elements are 0---i.e., $\bS^o$ and $\bS^m$ contain, respectively, $N^o$ and $N^m$ 1's in total. Moreover, the ones are located on different rows across the columns, which implies that the column vectors are linearly independent. The matrices $\bS^o$ and $\bS^m$ are therefore of full column rank. 

As a simple illustration, suppose $T=2, n=3$, and $y_{3,1}, y_{1,2}$ and $y_{3,2}$ are missing.
 Then, $\by^o = (y_{1,1}, y_{2,1}, y_{2,2})'$, $\by^m = (y_{3,1}, y_{1,2}, y_{3,2})'$ and 
\[
	\begin{bmatrix} y_{1,1} \\ y_{2,1} \\ y_{3,1} \\ y_{1,2} \\ y_{2,2} \\ y_{3,2} \end{bmatrix} 
	= \underbrace{
	\begin{bmatrix} 1 & 0 & 0 \\ 0 & 1 & 0 \\ 0 & 0 & 0 \\ 0 & 0 & 0 \\ 0 & 0 & 1 
	\\ 0 & 0 & 0 \end{bmatrix}}_{\bS^o}
	\begin{bmatrix} y_{1,1} \\ y_{2,1} \\ y_{2,2} \end{bmatrix}	
	+
	\underbrace{
	\begin{bmatrix} 0 & 0 & 0 \\ 0 & 0 & 0 \\ 1 & 0 & 0 \\ 0 & 1 & 0 \\ 
		0 & 0 & 0	\\ 0 & 0 & 1 \end{bmatrix}}_{\bS^m} 
	\begin{bmatrix} y_{3,1} \\ y_{1,2}  \\ y_{3,2} \end{bmatrix}. 
\]
For a second illustration, suppose $y_{1,t}$ is only observed every 3 periods at $t=3,6,9,\ldots,$ whereas $y_{2,t},\ldots,y_{n,t} $ are observed every period for $t=1,\ldots, T$. Then, $\bS^o$ is block-diagonal consisting of diagonal blocks $\bS_1^o, \bS_2^o,\ldots, \bS_T^o$, i.e., $\bS^o = \text{diag}(\bS_1^o, \bS_2^o,\ldots, \bS_T^o)$ and $\bS^m = \text{diag}(\bs_1^m, \bs_2^m,\ldots, \bs_T^m)$, where $\bS_t^o = \mathbf{I}_n$ and $\bs_t^m = \emptyset$ if $t$ is divisible by 3; otherwise
\[
	\bS_t^o = \begin{bmatrix}	\mathbf{0}_{n-1}' \\  \mathbf{I}_{n-1}\end{bmatrix},	              
	 \quad  
	\bs_t^m = \begin{bmatrix}	1 \\ \mathbf{0}_{n-1} \end{bmatrix}.
\]
In general, the selection matrices $\bS^o$ and $\bS^m$ are sparse and can be constructed easily even for complex missing data patterns.

Now, stacking~\eqref{eq:obs} over $t=1,\ldots, T$, and using the expression in~\eqref{eq:selection}, one can rewrite the model more compactly as
\begin{equation} \label{eq:obs_stacked}
	\bG^o\by^o + \bG^m\by^m  = \bW\valpha + \bX\vbeta + \vepsilon, \quad \vepsilon \sim\distn{N}(\mathbf{0}_{Tn},\vSigma),
\end{equation}
where  $\valpha = (\valpha_1',\ldots, \valpha_T')'$ and $\vSigma = \text{diag}(\vSigma_1,\ldots, \vSigma_T)$.\footnote{If the right-hand side of \eqref{eq:obs} does not contain any lagged values of $\by_t$, then $\bG^m = \bS^m$ and $\bG^o = \bS^o$. Otherwise, $\bG^o$ and $\bG^m$ become products of certain difference matrices and selection matrices, as illustrated in Example~\ref{ex:VAR}.} We assume $\bG^m$ has full column rank, which is satisfied for most commonly-used models (and can be easily verified in practice). Below we provide two examples to show how a dynamic factor model and a VAR($p$) can be expressed in the form of~\eqref{eq:obs_stacked}.

\begin{example}\label{ex:DFM}\rm Consider the following dynamic factor model with stochastic volatility:
\begin{align*}
	\by_t & = \bA_1\bff_t + \vepsilon_t, &\vepsilon_t&\sim\distn{N}(\mathbf{0}_n,\vSigma_t),  \\
	\bff_t & = \vPhi_1\bff_{t-1} + \cdots + \vPhi_q\bff_{t-q} + \vepsilon_t^{\bff}, 
	&	\vepsilon_t^{\bff} & \sim \distn{N}(\mathbf{0}_k, \vOmega_t),
\end{align*}
where $\vSigma_t = \text{diag}(\e^{h_{1,t}}, \ldots, \e^{h_{n,t}})$, $\vOmega_t = \text{diag}(\e^{h_{n+1,t}}, \ldots, \e^{h_{n+k,t}})$ are diagonal matrices with time-varying variances, and $\by_t$ is partitioned into two subvectors $\by_t^{o}$ and $\by_t^{m}$. Using the identity in \eqref{eq:selection}, the observation equation of this dynamic factor model can be expressed in the form in \eqref{eq:obs_stacked} as:
\begin{align*}
	\bS^o\by^{o} + \bS^{m}\by^{m} = \bH_{\bA_1}\bff + \vepsilon, \quad \vepsilon \sim\distn{N}(\mathbf{0}_{Tn},\vSigma),
\end{align*}
where $\bH_{\bA_1} = (\mathbf{I}_T\otimes \bA_1)$, $\vSigma = \text{diag}(\vSigma_1,\ldots, \vSigma_T)$ and $\otimes$ denotes the Kronecker product. One can consider a more general dynamic factor model in which the observation equation contains lagged values of the dynamic factors, say, $\bff_{t-1}, \ldots, \bff_{t-p}$. In this case one can simply redefine the matrix $\bH_{\bA_1}$ to include them in the observation equation.
\end{example}

\begin{example} \label{ex:VAR} \rm The next example is a VAR($p$) with an outlier component:
\begin{equation}\label{eq:var}
	\by_{t} = \mathbf{b}_0 + \bB_{1}\by_{t-1} + \bB_{2}\by_{t-2}
	+ \cdots + \bB_{p}\by_{t-p} + \vepsilon_{t},\quad \vepsilon_{t}\sim \distn{N}(\mathbf{0}_n,\vSigma_t),
\end{equation}
where $\vSigma_t = o_t^2\bQ$, $\bQ$ is a covariance matrix and $o_t$ follows a 2-part distribution with a point mass at 1 and a uniform distribution on the interval $(2,10)$. Then, stacking~\eqref{eq:var} over $t=1,\ldots, T$, we obtain
\begin{equation}\label{eq:VAR_stacked}
	\bH_{\bB}\by = \bc_{\bB} + \vepsilon, \quad \vepsilon\sim \distn{N}(\mathbf{0}_{Tn}, \vSigma),
\end{equation}
where 
\begin{equation}\label{eq:H}
\resizebox{.9\hsize}{!}{$
\bc_{\bB} = \begin{bmatrix}
\mathbf{b}_0  +\sum_{j=1}^{p}\bB_{j}\by_{1-j}\\
\mathbf{b}_0  +\sum_{j=2}^{p}\bB_{j}\by_{2-j}\\
\vdots\\
\mathbf{b}_0 + \bB_{p}\by_0\\
\mathbf{b}_0  \\
\vdots \\
\mathbf{b}_0 
\end{bmatrix}, \;
\bH_{\bB} = \begin{bmatrix}
\mathbf{I}_n & \mathbf{0}_{n\times n} & \cdots & \cdots & \cdots & \cdots& \cdots & \mathbf{0}_{n\times n}\\
-\bB_1 & \mathbf{I}_n & \mathbf{0}_{n\times n} & \cdots &\cdots & \cdots & \cdots & \mathbf{0}_{n\times n} \\
-\bB_2 & -\bB_1 &\mathbf{I}_n & \mathbf{0}_{n\times n} & \cdots &  &  & \mathbf{0}_{n\times n}\\
\vdots & \ddots & \ddots & \ddots & \ddots & \ddots &  & \vdots \\
-\bB_p & \cdots &  & -\bB_1 & \mathbf{I}_n & \mathbf{0}_{n\times n} &   & \vdots\\
\mathbf{0}_{n\times n} &  &  &  & \ddots & \ddots & \ddots & \vdots\\
\vdots &  & \ddots &  &  \ddots & \ddots & \ddots & \vdots\\
\mathbf{0}_{n\times n} & \cdots & \mathbf{0}_{n\times n} & -\bB_{p} &  \cdots & -\bB_2 & -\bB_{1} & \mathbf{I}_n
\end{bmatrix}.$
}
\end{equation}
Again, using the identity in \eqref{eq:selection}, we obtain
\begin{align*}
	\bH_{\bB}\bS^o\by^{o} + \bH_{\bB}\bS^{m}\by^{m} = \bc_{\bB} + \vepsilon, \quad \vepsilon \sim\distn{N}(\mathbf{0}_{Tn},\vSigma),
\end{align*}
which is in the form of \eqref{eq:obs_stacked} with $\bG^o = \bH_{\bB}\bS^o$, $\bG^m = \bH_{\bB}\bS^m$, $\bX = \mathbf{I}_{Tn}$ and $\vbeta = \bc_{\bB}$.\footnote{For notational convenience, in the derivation we condition on the initial conditions $\by_{0},\ldots, \by_{1-p}$. These initial conditions could potentially have missing data, but they can be sampled in a separate step. Since $p$ is much smaller than $T$ in most applications, this extra step is typically computationally trivial. Alternatively, one can jointly sample the missing data in the initial conditions and the sample by redefining $\by$ and the associated matrices.}

\end{example}

Using the expression in~\eqref{eq:obs_stacked}, next we derive the conditional distribution of $\by^{m}$ given $\by^{o}$ and other model parameters and latent variables, which we collectively denote as $\vtheta$. Intuitively, since the joint distribution of $(\by^{o}, \by^{m})$ is Gaussian conditional on $\vtheta$, the conditional distribution of $\by^{m}$ given $\by^{o}$ and $\vtheta$ is also Gaussian by the properties of the Gaussian distribution. More precisely, it follows from \eqref{eq:obs_stacked} that $p(\by^{m}\gvn\by^o,\vtheta)$ can be expressed as
\begin{align*}
	p(\by^{m}\gvn\by^o,\vtheta) \propto & \exp\left\{-\frac{1}{2}\left(\bG^m\by^m + \bG^o\by^o - \bW\valpha - \bX\vbeta\right)'\vSigma^{-1}\left(\bG^m\by^m + \bG^o\by^o - \bW\valpha - \bX\vbeta\right)\right\} \\
\propto & \exp\left\{-\frac{1}{2}\left[\by^{m \prime}\bG^{m \prime}\vSigma^{-1}\bG^m\by^m
 -2\by^{m \prime}\bG^{m \prime}\vSigma^{-1}(\bW\valpha + \bX\vbeta - \bG^o\by^o)\right]\right\}.
\end{align*}
Next, let $\bK_{\by^m} = \bG^{m \prime}\vSigma^{-1}\bG^m$, which is an $N_m\times N_m$ non-singular matrix---since $\bG^m$  has full column rank---and is thus invertible. Furthermore, let $\vmu_{\by^m} = \bK_{\by^m}^{-1}\bG^{m \prime}\vSigma^{-1}(\bW\valpha + \bX\vbeta - \bG^o\by^o).$ Then, by completing the square in $\by^m$, one can write the conditional distribution of $\by^m$ as 
\begin{align*}
	p(\by^{m}\gvn\by^o,\vtheta) \propto  &\exp\left\{-\frac{1}{2}\left(\by^{m \prime}\bK_{\by^m}\by^m
	 -2\by^{m \prime}\bK_{\by^m}\vmu_{\by^{m}}\right)\right\} \\
	\propto & \exp\left\{-\frac{1}{2} \left(\by^m-\vmu_{\by^m}\right)'\bK_{\by^m}\left(\by^m-\vmu_{\by^m}\right)\right\}.
\end{align*}
Thus, we have shown that the joint conditional distribution of the missing data given the observed data is Gaussian with mean vector $\vmu_{\by^m}$ and precision matrix $\bK_{\by^m}$:
\begin{equation} \label{eq:ym_cond}
	(\by^{m}\gvn\by^o,\vtheta)\sim 
	\distn{N}\left(\vmu_{\by^m},\bK_{\by^m}^{-1}\right).
\end{equation}
Since both $\bG^m$ and $\vSigma$ are band matrices, so is the precision matrix 
$\bK_{\by^m}$. Therefore, we can use the precision-based sampler of \citet{CJ09} to draw 
$\by^{m}$ efficiently. We summarize the sampler in Algorithm~\ref{alg:precision}.

\begin{algorithm}[H]
\caption{Sampling $(\by^{m}\gvn\by^o,\vtheta)\sim \distn{N}\left(\vmu_{\by^m},\bK_{\by^m}^{-1}\right)$.}
\label{alg:precision}
Given $\vmu_{\by^{m}}$ and $\bK_{\by^m}$, complete the following steps.
\begin{enumerate}
	\item Obtain the Cholesky factor $\bC$ of $\bK_{\by^m}$ such that $\bK_{\by^m} = \bC\bC'$.
	
	\item Solve $\bC'\bv = \bx$ for $\bv$ by backward substitution, where $\bx\sim\distn{N}(\mathbf{0}_{N^m},\mathbf{I}_{N^m})$.	
		
	\item Return $\by^{m} = \vmu_{\by^{m}} + \bv$.
	
\end{enumerate}
\end{algorithm}

This paper focuses on Bayesian estimation using MCMC methods. But the above results are also useful for other estimation methods. For example, the conditional distribution in~\eqref{eq:ym_cond} can be used in conjunction with the expectation-maximization algorithm to obtain the maximum likelihood estimate of $\vtheta$. Alternatively, one can directly maximize the observed-data likelihood, which can be evaluated using the identity $p(\by^o\gvn\vtheta) = p(\by^o, \by^m \gvn\vtheta) / p(\by^m \gvn \by^o, \vtheta)$. The above results show that both densities on the right-hand side are Gaussian and can be evaluated quickly.

The setup so far is suitable for applications with missing data patterns such as unbalanced panels and ragged edges. In many situations, however, the researcher has additional information on the missing data. A prominent example is a mixed-frequency model in which the high-frequency observations of the low-frequency flow variables are treated as missing data, and these missing observations are linked to multiple low-frequency observations. Next, we show how one can incorporate additional information to update the conditional distribution of the missing data.

\subsection{Conditioning on Additional Information}

The previous section discussed how one can efficiently sample the vector of missing data $\by^{m}$  conditional on the observed data $\by^{o}$ and the model parameters $\vtheta$. Additional information is available in many applications, and it is often desirable or necessary to incorporate the new information into the analysis. For example, in a mixed-frequency setting with both monthly and quarterly variables, a common approach is to treat the monthly observations of the quarterly flow variables as missing, and these missing values are then mapped to the observed values via some inter-temporal constraints. Another example is ragged edge settings where the latest values of some variables are not yet released, but high-quality nowcasts (e.g., from surveys of professional forecasters) are available. Below we show how one can modify the proposed sampling approach to handle various settings with additional information.

To keep the proposed framework general that can handle a wide variety of information settings, suppose there is an additional $M\times 1$ vector of observables, say, $\bz$, that is available for sharpening the inference on $\by^{m}$. We consider two types of mappings that connect $\bz$ to $\by^{m}$. In the first case, suppose $\bz$ can be mapped to the missing data exactly via the linear system:
\begin{equation} \label{eq:hard}
	\bz = \bM \by^{m},
\end{equation}
where $\bM$ is an $M\times N^m$ matrix specifying the $M$ exact linear relationships. We refer to this type of additional information as hard constraints. One important example is the inter-temporal constraints for mixed-frequency settings based on a log-linear approximation proposed in \citet{MM03, MM10}. More specifically, suppose $y_{i,t}^{m}$ is the missing monthly value of the $i$-th variable at month $t$. Let $z_{i,t/3}$ denote the corresponding observed quarterly value (note that $z_{i,t/3}$ is only observed for every third month). Then, a standard log-linear approximation to an arithmetic average of the quarterly variable can be expressed as:
\begin{equation} \label{eq:aggre}
	z_{i,t/3} = \frac{1}{3}y_{i,t}^{m}+\frac{2}{3}y_{i,t-1}^{m} + y_{i,t-2}^{m}
		+ \frac{2}{3}y_{i,t-3}^{m}+\frac{1}{3}y_{i,t-4}^{m}
\end{equation}
for $t=3,6,9,\ldots$. By stacking \eqref{eq:aggre} and defining $\bM$ appropriately, the exact linear restrictions in \eqref{eq:aggre} can be written in the form in \eqref{eq:hard}. For balanced monthly and quarterly variables, $M = N^m/3$.

Even though the mapping considered in \eqref{eq:aggre} is technically based on a log-linear approximation, in most applied work it is treated as an exact linear relationship. A more appropriate approach might be to explicitly allow for measurement or approximation errors. In addition, there are other situations where allowing for measurement errors is appropriate (e.g., mapping nowcasts from professional forecasters to the underlying endogenous variables). Hence, we consider an alternative mapping that includes measurement errors of the form:
\begin{equation} \label{eq:soft}
	\bz = \bM \by^{m} + \vepsilon^z, \quad \vepsilon^z\sim\distn{N}(\mathbf{0}_M,\bO),
\end{equation}
where $\bO$ is a fixed diagonal covariance matrix that encodes the magnitude of the measurement errors. We refer to this type of additional information as soft constraints.

After providing a general setting to incorporate additional information, next we discuss how the sampling of the missing data can be modified given this new information set. First, we consider the case of hard constraints. Recall that the missing data conditional only on the observed data and model parameter is Gaussian as specified in \eqref{eq:ym_cond}. Therefore, sampling the missing data conditioning on the exact linear restrictions in \eqref{eq:hard} amounts to drawing from the degenerate Gaussian distribution $\distn{N}\left(\vmu_{\by^m},\bK_{\by^m}^{-1}\right)1(\bM \by^{m} = \bz)$, where $1(\cdot)$ is the indicator function. There are efficient algorithms that can be used to sample from $\distn{N}\left(\vmu_{\by^m},\bK_{\by^m}^{-1}\right)$ so that $\bM \by^{m} = \bz$, such as Algorithm 2.6 in \citet{RH05} and Algorithm~2 in \citet{CCZ17}. In particular, we can first sample $\bu \sim \distn{N}\left(\vmu_{\by^{m}},\bK_{\by^m}^{-1}\right)$ using Algorithm~\ref{alg:precision}. Then, we update the condition set augmented with $\bz = \bM \by^{m}$ by computing 
\[
	\by^{m} = \bu + \bK_{\by^m}^{-1} \bM'(\bM\bK_{\by^m}^{-1}\bM')^{-1}(\bz -\bM\bu).
\]
It can be shown that $\by^{m}$ has the distribution $(\by^m\gvn\by^o,\vtheta, \bM\by^{m} = \bz)$. Algorithm~\ref{alg:gaulr} describes an efficient implementation in \citet{RH05} that avoids explicitly computing the inverse of $\bK_{\by^m}$ or $\bM\bK_{\by^m}^{-1}\bM'$. Using this implementation, the additional computational cost for conditioning on $\bz = \bM \by^{m}$ is relatively low for $M \ll N^m$. For large $M$, this algorithm would involve a few large, dense matrices, and the computations could be more intensive. 

\begin{algorithm}[H]
\caption{Sampling $(\by^{m}\gvn\by^o,\vtheta, \bz = \bM\by^m)$ with hard constraints, where $(\by^{m}\gvn\by^o,\vtheta) \sim \distn{N}\left(\vmu_{\by^{m}},\bK_{\by^m}^{-1}\right)$.}
\label{alg:gaulr}
Given the parameters $\vmu_{\by^m}$ and $\bK_{\by^m}$, complete the following steps.
\begin{enumerate}
	\item Obtain the Cholesky factor $\bC$ of $\bK_{\by^m}$ such that $\bK_{\by^m}=\bC\bC'$.
	
	\item Sample $\bu\sim\distn{N}\left(\vmu_{\by^{m}},\bK_{\by^m}^{-1}\right)$ using Algorithm~\ref{alg:precision}.		
		
	\item Solve $\bC\bC'\bU = \bM'$ for $\bU$ by forward and backward substitution.
	
	\item Solve $\bM\bU \bV =\bU'$ for $\bV$.
		
	\item Return $\by^m = \bu + \bV'(\bz - \bM\bu)$.
	
\end{enumerate}
\end{algorithm}

Next, we consider the case of soft constraints. Essentially, we update the conditional distribution of the missing data $\by^m$ given the new information specified in~\eqref{eq:soft}. Therefore, one can view the original Gaussian distribution of $\by^m$ in \eqref{eq:ym_cond} as the `prior distribution' and the new information in \eqref{eq:soft} as the `likelihood'. Then, by standard Bayesian updating, we obtain
\begin{equation} \label{eq:ym_cond_soft}
	(\by^m \gvn \by^{o}, \vtheta, \bz)\sim 
	\distn{N}\left(\overline{\vmu}_{\by^m}, \overline{\bK}_{\by^m}^{-1}\right),
\end{equation} 
where
\[
	\overline{\bK}_{\by^m} = \bM'\bO^{-1}\bM + \bK_{\by^m}, \quad
	\overline{\vmu}_{\by^m} = \overline{\bK}_{\by^m}^{-1}\left(\bM'\bO^{-1}\bz + \bK_{\by^m}\vmu_{\by^m}\right).
\]
Since for most applications the matrices $\bM, \bO$ and $\bK_{\by^m}$ are all banded, so is the precision matrix $\overline{\bK}_{\by^m}$. Hence, the precision-based sampler in Algorithm~\ref{alg:precision} can be directly applied to sample $\by^m$ efficiently; we simply replace $\bK_{\by^m}$ and $\vmu_{\by^m} $ by $\overline{\bK}_{\by^m}$ and $ \overline{\vmu}_{\by^m} $, respectively.

Compared to Algorithm~\ref{alg:gaulr} for the case of hard constraints, sampling from \eqref{eq:ym_cond_soft} is much faster and scales well to high-dimensional settings. For approximate inter-temporal restrictions such as \citet{MM03, MM10}, the latter sampler is naturally preferable. For other exact inter-temporal restrictions, empirically one can approximate these hard constraints by setting the diagonal elements of $\bO$ to be very small (e.g., $10^{-8}$).

\section{A Monte Carlo Study} \label{s:MC}

In this section we conduct a series of Monte Carlo experiments to assess the speed and accuracy of the proposed precision-based methods for drawing the latent missing observations relative to Kalman-filter based methods. In the first subsection we consider mixed-frequency settings in which the missing data are the high-frequency observations of the low-frequency variables. We then consider unbalanced panels in the following subsection.

All datasets are generated from the following VAR: 
\[
	\by_{t} = \mathbf{b}_0 + \bB_{1}\by_{t-1} + \bB_{2}\by_{t-2} + \cdots + \bB_p\by_{t-p}  +
		\vepsilon_{t}, \quad \vepsilon_{t}\sim \distn{N}(\mathbf{0}_n,\vSigma),
\]
where $\by_{t}=(\by_{t}^{o \prime},\by_{t}^{m \prime})'$ is an $n\times1$ vector of mixed-frequency data, $\by_{t}^{o}$ is an $n^{o}\times 1$ vector of (observed) high-frequency variables and $\by_{t}^{m}$ is an $n^m\times 1$ vector of (missing) high-frequency observations of the low-frequency variables. In addition, low-frequency variables $z_{i,t/3}, i=1,\ldots,n^m,$ are observed at $t=3,6,9,\ldots$, which can be used to inform the values of the missing $\by_{t}^{m}$ via \eqref{eq:hard} or \eqref{eq:soft}. For the baseline case we set $p=5$. Furthermore, we generate the model parameters as follows. We set $\mathbf{b}_0 = 0.01\times \mathbf{1}_{n}$. The diagonal elements of the first VAR coefficient matrix are iid uniform $\distn{U}(0, 0.5)$ and the off-diagonal elements are $\distn{U}(-0.2, 0.2)$. All elements of the higher VAR coefficient matrix are iid $\distn{N}(0, 0.05^2/l^2)$, where $l$ is the lag length. The error covariance matrix $\vSigma$ is generated from the inverse-Wishtart distribution $\distn{IW}(n+10, 0.07\mathbf{I}_n + 0.03\mathbf{1}_n\mathbf{1}_n')$.

For each simulated dataset $r=1,\ldots, R$, we estimate the missing observations $\by_{t}^{m}$ using 4 methods: the precision-based sampler with the hard inter-temporal constraints in \eqref{eq:hard}, the precision-based sampler with the soft constraints in \eqref{eq:soft}, the simulation smoother of \citet{CK94} as implemented in the code provided by \citet{SS15}, and the simulation smoother of \citet{DK02}.\footnote{The implementation in \citet{SS15} uses a compact state-space representation to draw the missing observations. This representation removes the monthly observations from the state vector that appears in the measurement equation. Consequently, it reduces the dimension of the state vector and is generally more efficient. In contrast, the simulation smoother of \citet{DK02} requires the model in a standard companion form (see Appendix~B for details) that results in a higher dimensional state vector.} For both the simulation smoothers of \citet{CK94} and \citet{DK02}, we impose the hard constraints. For the precision-based sampler with soft constraints, we set the diagonal elements of the measurement error covariance matrix $\bO$ to be $10^{-8}$. Hence, we view it as an approximation of the hard constraints so that results from the 4 methods are comparable. 

Finally, we implement a standard normal prior for the VAR coefficients and an inverse-Wishart prior for the error covariance matrix. More specifically, let $\vbeta=\text{vec}\left(\left[\mathbf{b}_{0},\mathbf{B}_{1},\ldots,\mathbf{B}_{p}\right]'\right)$ denote the $k\times 1$ vector of all VAR coefficients with $k=n(np+1)$. Then, the priors are $\vbeta \sim\mathcal{N}(\mathbf{0}_{k},\mathbf{I}_{k})$ and $\vSigma\sim\mathcal{IW}(5,\mathbf{I}_{n})$.

\subsection{Missing Observations of Low-Frequency Variables}

We consider DGPs of different dimensions with $T=300$: small ($n=6, n^{o}=5, n^{m}=~1)$, medium ($n=11, n^{o}=10, n^{m}=1)$ and large ($n=16, n^{o}=15, n^{m}=1)$. We also investigate settings with a larger number of unobserved variables with $n^{m} = 5$. For each design, we generate $R=100$ datasets from the mixed-frequency VAR as described above. We then fit each dataset using a Gibbs sampler that draws sequentially the missing observations and model parameters. In particular, we use 4 different methods to sample the missing observations. To assess the accuracy of the different methods, we compute the mean squared error (MSE) of the estimated missing observations against the actual values. More specifically, for each dataset with missing observations $\by^{m (i)}_1, \ldots, \by^{m (i)}_T$, $i=1,\ldots, R,$, and each method with posterior mean vector $\hat{\by}^{m (i,j)}, j=1,\ldots, 4$, we compute 
$\text{MSE}_i(\hat{\by}^{m (i,j)}) = \sum_{t=1}^T||\by_t^{m (i)} - \hat{\by}_t^{m (i,j)}||^2/T$, where $||\cdot||$ is the $\ell^2$-norm. We further summarize the accuracy by averaging over the $R$ MSEs for each design, and the results are reported in Table~\ref{tab:MSE}.

We also report the computation time, based on 15,000 MCMC draws with a burn-in period of 5,000 draws.\footnote{The computation time is based on a standard desktop with an Intel Xeon W-2223 @ 3.6GHz processor and 16 GB of RAM and the code is implemented in $\matlab$.} Since all four methods aim to draw from the same distribution---namely, the conditional distribution of the missing observations given the observed data and model parameters---in principle they should give the same MSEs. Indeed, they give very similar MSEs in our simulations; the small discrepancies are mainly due to numerical and simulation errors. In terms of runtime, it is clear that the proposed precision-based methods are more computationally efficient compared to the Kalman-filter based method across a range of settings. 

\begin{table}[H]
\caption{Mean squared errors of the estimated missing observations and computation time using four methods: the proposed precision-based method with hard inter-temporal constraints (P-hard), the precision-based method with soft inter-temporal constraints (P-soft), the simulation smoother of \citet{CK94} implemented in \citet{SS15} (CK) and the simulation smoother of \citet{DK02} (DK).}
\label{tab:MSE} \centering 
\begin{tabular}{cccccccccc}
\hline 
 &  & \multicolumn{4}{c}{MSE} & \multicolumn{4}{c}{Computation time (minutes)}\tabularnewline
$n^{m}$  & $n^{o}$  & P-hard  & P-soft  & DK & CK & P-hard  & P-soft  & DK  & CK\tabularnewline
\hline 
1  & 5  & 0.004 & 0.004 & 0.004 & 0.005 & 0.7 & 0.6 & 7 & 5\tabularnewline
\rowcolor{lightgray} 1  & 10  & 0.004 & 0.004 & 0.004 & 0.004 & 3 & 3 & 31 & 9\tabularnewline
1  & 15 & 0.004 & 0.004 & 0.004 & 0.004 & 13 & 13 & 61 & 16\tabularnewline
\rowcolor{lightgray} 5  & 5 & 0.005 & 0.005 & 0.005 & 0.005 & 8 & 4 & 23 & 24\tabularnewline
5  & 10  & 0.005 & 0.005 & 0.005 & 0.005 & 16 & 12 & 51 & 35\tabularnewline
\rowcolor{lightgray} 5  & 15 & 0.004 & 0.004 & 0.004 & 0.005 & 38 & 35 & 106 & 51\tabularnewline
\hline 
\end{tabular}
\end{table}

Table~\ref{tab:MSE} reports the runtime of full MCMC estimation. When the dimension of the VAR increases, the computational complexity of simulating the VAR coefficients dominates, and it might give the impression that the runtime of the four methods converges. To better understand how the proposed methods perform across a wider range of settings, next we compare only the runtime of sampling the missing observations. 

\begin{figure}[H]
    \centering
   \includegraphics[width=.95\textwidth]{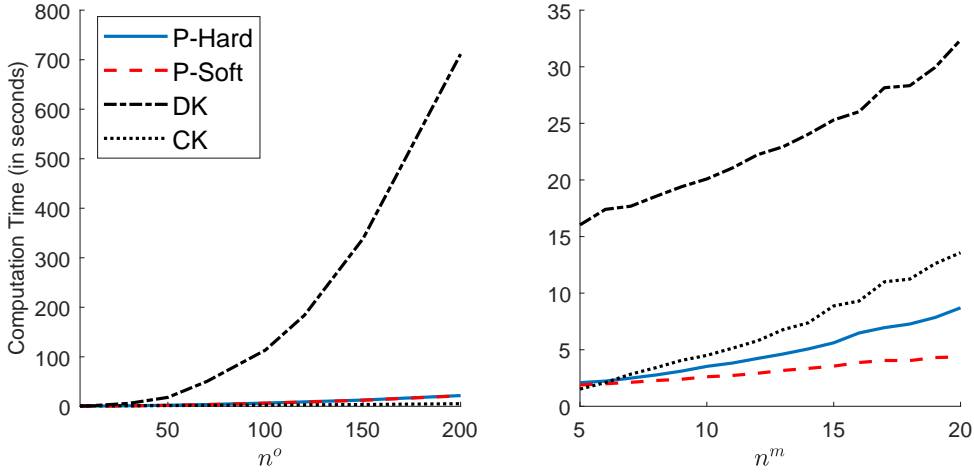}
   \caption{Computation time of obtaining 10 draws against $n^{o}$ and $n^{m}$, the numbers of observed and partially unobserved variables, respectively, with $T=300$ and $p=5$. The four methods are: precision-based sampler with hard inter-temporal constraints (P-hard), precision-based sampler with soft constraints (P-soft), the simulation smoother of \citet{CK94} implemented in \citet{SS15} (CK) and the simulation smoother of \citet{DK02}.}
   \label{fig:simulation_n}	
\end{figure}

First, Figure~\ref{fig:simulation_n} reports the runtime of sampling ten draws of the missing observations using the four methods for a range of $n^{o}$ and $n^{m}$. It is clear that both precision-based methods compare favorably to the Kalman-filter based methods, and both scale well to high dimensional settings. In addition, the variant with soft restrictions is especially efficient when there are a large number of variables with missing observations. 

\begin{figure}[H]
    \centering
   \includegraphics[width=.95\textwidth]{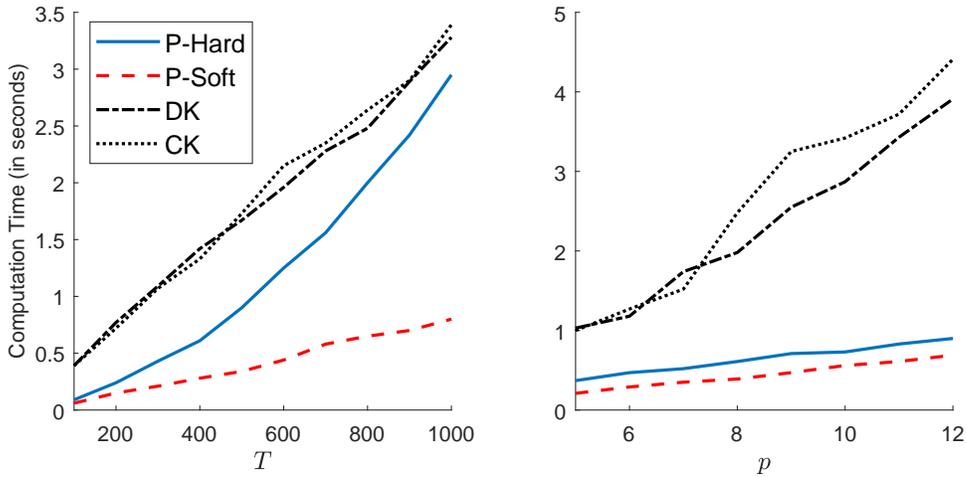}
   \caption{Computation time of obtaining 10 draws against $T$ and $p$, the numbers of time periods and lags, respectively, with $n^{m}=5$ and $n^{o}=10$. The four methods are: precision-based sampler with hard inter-temporal constraints (P-hard), precision-based sampler with soft constraints (P-soft), the simulation smoother of \citet{CK94} implemented in \citet{SS15} (CK) and the simulation smoother of \citet{DK02}.}
   \label{fig:simulation_pT}	
\end{figure}

Next, Figure~\ref{fig:simulation_pT} reports the runtime of sampling ten draws of the missing observations for a range of sample sizes $T$ and lag lengths $p$. While both precision-based methods perform well, the version with soft constrains does substantially better and scales well to very large $T$ and $p$. It is also worth mentioning that to apply the Kalman filter, one needs to redefine the states so that the observation equation depends only on the current (redefined) state. When $p$ is large, the dimension of this new state vector is large. That is one reason why the Kalman-filter based methods become more computationally intensive when $p$ is large. In contrast, the computational costs of the precision-based methods remain low even for long lag lengths.

\subsection{Unbalanced Panels} \label{ss:unbalanced}

In this section we illustrate the performance of the proposed precision-based methods in settings involving unbalanced panels where different variables are missing in different time periods. To that end, we conduct a series of simulations using a medium-size VAR ($n=13,$ $n^{o}=10$, $n^{m}=3)$ with a sample size of $T=300$. We simulate the data using the same DGP as described above, but here we assume that
the low-frequency variables are also missing for selected time periods (i.e., in addition to their missing high-frequency observations). More specifically, the first low-frequency variable is completely missing for the first 30 periods; the second is missing from $t=150,\ldots, 180$; and the third is missing for the last 30 periods. 

\begin{table}[H]
 \centering
\caption{Mean squared errors of the estimated missing observations and computation time averaging over $R=100$ replications using four methods: the proposed precision-based method with hard inter-temporal constraints (P-hard), the precision-based method with soft inter-temporal constraints (P-soft), the simulation smoother of \citet{CK94} implemented in \citet{SS15} (CK) and the simulation smoother of \citet{DK02} (DK).}
\label{tab:MSEunbalanced}
\begin{tabular}{cccccccccc}
\hline 
 &  & \multicolumn{4}{c}{MSE} & \multicolumn{4}{c}{Computation time (minutes)}\tabularnewline
$n^{m}$ & $n^{o}$ & P-hard & P-soft & DK & CK & P-hard & P-soft & DK & CK\tabularnewline
\hline 
3 & 10 & 0.004 & 0.004 & 0.004 & 0.009 & 8 & 7 & 40 & 14\tabularnewline
\hline 
\end{tabular}
\end{table}

Table~\ref{tab:MSEunbalanced} reports the MSEs and computation time for estimating the missing observations of the three low-frequency variables. Similar to the previous simulated experiments, here we also find that the MSEs across the four methods are very similar. As before, the proposed precision-based methods are computationally more efficient than both Kalman-filter based simulation smoothers.

\begin{figure}[H]
\includegraphics[width=1\textwidth]{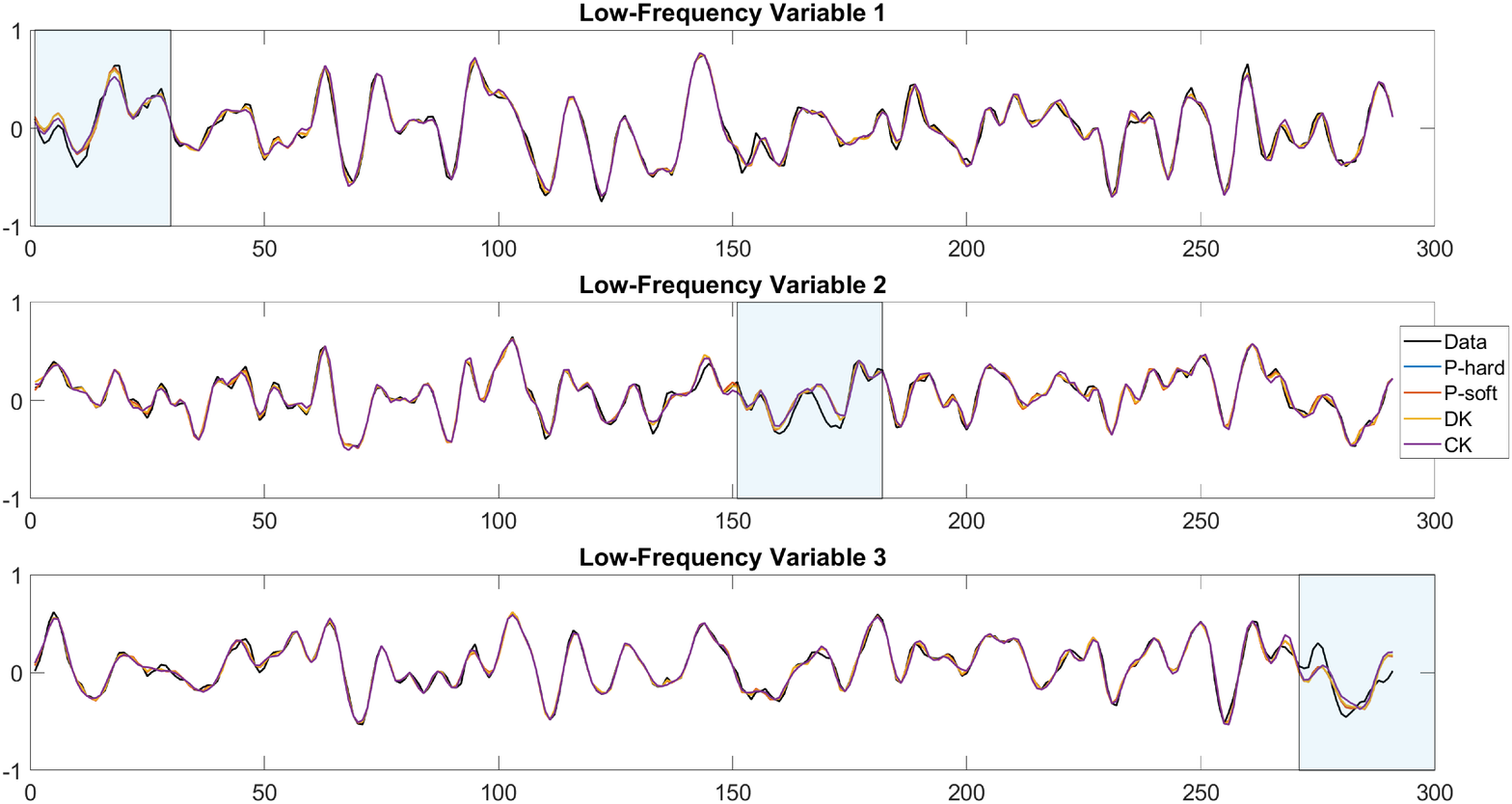}
\caption{Posterior means of the low-frequency variables obtained using four methods: the proposed precision-based method with hard inter-temporal constraints (P-hard), the precision-based method with soft inter-temporal constraints (P-soft), the simulation smoother of \citet{CK94} implemented in \citet{SS15} (CK) and the simulation smoother of \citet{DK02} (DK). The box in each panel highlights the time periods in which the low-frequency variable is completely missing.}
\label{fig:unbalancedfig}
\end{figure}

Next, Figure~\ref{fig:unbalancedfig} plots the posterior means of the missing observations of the low-frequency variables obtained using the four methods against the actual simulated data. In the figure we highlight the time periods in which each low-frequency variable is completely missing. All four methods produce very similar posterior estimates of the missing observations during these periods as expected. We therefore conclude that the precision-based methods can handle any arbitrary missing data pattern as well as standard filtering and smoothing methods, but they are more computationally efficient.

\section{Empirical Applications} \label{s:applications}
 
We demonstrate the proposed precision-based approach via two empirical applications involving two popular large-scale models and widely different missing data patterns. In the first application, we use a large mixed-frequency VAR with stochastic volatility to generate weekly estimates of real GDP. In the second application, we extract latent factors from unbalanced datasets using a dynamic factor model with stochastic volatility.

\subsection{A Weekly State-Space Mixed-Frequency VAR}

In the first application we illustrate how the proposed precision-based approach can be used to estimate a Bayesian VAR with 22 variables in weekly, monthly and quarterly frequencies. Following the seminal work by \citet{SS15}, we model all variables at the highest observed frequency, and treat the high-frequency observations of the low-frequency variables as missing data. One key advantage of this approach is that it produces interpolated (historical) estimates of the low-frequency variables at a higher frequency. For example, \citet{SS15} use monthly and quarterly variables to obtain monthly GDP estimates, which are required as inputs for a variety of applications. In addition, this approach can also deliver more timely nowcasts by incorporating information in higher-frequency variables. Recent applications of this approach include \citet{BBJ19} and \citet{KMMP20, KMMP22}. 

We extend this line of work by including weekly macroeconomic and financial variables. By modeling all variables in weekly frequency, we are able to obtain weekly GDP estimates. Fitting a large mixed-frequency VAR, however, is computationally intensive as there are a large number of missing observations. Moreover, since the missing data pattern is irregular (e.g., each quarter or month does not always have the same number of weeks), the implementation of conventional methods is also more complex. In contrast, the proposed method can easily handle the irregular missing data pattern and it scales well to high dimensions.

The US dataset consists of 16 weekly variables (including raw steel production, retail sales, initial claims for unemployment benefits and various financial variables), 5 monthly variables (such as industrial production, CPI and labor market variables) and 1 quarterly variable (real GDP) from January 2013 to August 2022. Seven of the weekly variables are obtained from \citet{LMST22}, which they use to construct their Weekly Economic Index (WEI); other variables are sourced from the FRED database at the Federal Reserve Bank of St. Louis. More details about the data can be found in Appendix~A. 

Since our sample includes the COVID-19 pandemic, it is empirically important to allow for some form of heteroskedasticity or non-Gaussian errors, as demonstrated in recent papers such as \citet{Hartwig21}, \citet{LP22} and \citet{CCMM22}. We therefore incorporate the common stochastic volatility of \citet{CCM16} into a mixed-frequency VAR as follows:
\[
	\by_{t} = \mathbf{b}_0 + \bB_{1}\by_{t-1} + \bB_{2}\by_{t-2} + \cdots + \bB_p\by_{t-p}  +
		\vepsilon_{t}, \quad \vepsilon_t \sim\distn{N}(\mathbf{0},\e^{h_t}\vSigma),
\]
where $\by_{t}=(\by_{t}^{o \prime},\by_{t}^{m \prime})'$ is an $n\times1$ vector of mixed-frequency data, and $\by_{t}^{o}$ and $\by_{t}^{m}$ are the vectors of observed and missing variables, respectively. Note that the error covariance matrix is scaled by the common log-volatility $h_t$, which is modeled as a stationary AR(1) process:
\[
	h_t = \phi h_{t-1} + u_t^h, \quad u_t^h\sim\distn{N}(0,\sigma^2_h),
\]
for $t=2,\ldots, T$, where $|\phi|<1$ and the initial condition is specified as 
$h_{1}\sim\distn{N}(0,\sigma^2_h/(1-\phi^2))$. Here the unconditional mean of the AR(1) process is assumed to be zero for identification.\footnote{In preliminary work we also implemented a version of the model with the more flexible Cholesky stochastic volatility of \cite{CS05} and \citet{CCM19}, in which each variable has its own stochastic volatility process. However, given the large number of missing observations, we found that it was hard to pin down some of the stochastic volatility processes in simulations. In contrast, the common stochastic volatility worked well. An additional advantage of the common stochastic volatility is that it is order-invariant, whereas the Cholesky stochastic volatility is not.} 

Next, we describe the priors on the model parameters $\vbeta = \text{vec}([\mathbf{b}_0, \bB_1, \ldots, \bB_p]')$, $\vSigma, \phi$ and $\sigma^2_h$. Specifically, we assume the following independent priors:
\[
	\vbeta \sim\distn{N}(\vbeta_{0},\bV_{\vbeta}), \; \vSigma\sim\distn{IW}(\nu_0,\bS_0), \; \phi\sim \distn{N}(\phi_{0},V_{\phi})1(|\phi|<1),\; 	\sigma^2_h \sim \distn{IG}(\eta_{0},S_{0}),
\]	
where $1(\cdot)$ denotes the indicator function. The prior mean vector and covariance matrix of the VAR coefficients, $\vbeta_{0}$ and $\bV_{\vbeta}$ respectively, are chosen in the spirit of the Minnesota prior pioneered by \citet{DLS84} and \citet{litterman86}. More specifically, we set $\vbeta_{0} = \mathbf{0}_k$ to shrink the coefficients to zero. For the prior covariance matrix $\bV_{\vbeta}$, it is assumed to be diagonal such as
\begin{align*}
	\text{Var}(B_{l,ii}) & = \frac{\kappa_1}{l^2}, \; l = 1,\ldots, p, \; i = 1,\ldots, n, \\
	\text{Var}(B_{l,ij}) & = \frac{\kappa_2 s_i^2}{l^2 s_j^2},\; l = 1,\ldots, p,\; i,j = 1,\ldots, n, \; i\neq j, \\
	\text{Var}(b_{0,i})	& = 100 s_i^2, \; i = 1,\ldots, n,	
\end{align*}
where $B_{l,ij}$ is the $(i,j)$ element of $\bB_l$, $s_r^2$ denotes the sample variance of the residuals from an AR(4) model for the variable $r$ for $r=1,\ldots, n$. To implement cross-variable shrinkage, i.e., shrinking coefficients on lags of other variables more strongly to zero than on own lags, we set $\kappa_1 = 0.04$ and $\kappa_2 = 0.01$. Finally, we set $\nu_0 = n+3, \bS_0 = \mathbf{I}_n$, $\eta_{0} = 10$ and $ S_{0} = 0.004$ so that the prior means of $\vSigma$ and $\sigma^2_h$ are $0.5\mathbf{I}_n$ and $0.021^2$.

We model all variables in weekly frequency, and the missing observations of the monthly or quarterly variables are linked to their corresponding observed values via inter-temporal constraints similar to those in \citet{MM03,MM10}. Compared to standard mixed-frequency settings involving only quarterly and monthly variables, the inter-temporal constraints here are more complex, as there might be different numbers of weeks in different months or quarters. More specifically, given the releasing date $t$ of a monthly/quarterly variable, let $n_{i,t}^{w}$ denote the number of weeks between $t$ and the last releasing date. Then, each observed monthly/quarterly variable $z_{i,t}$ is linked to the missing observations $y_{i,t}^m$ via the inter-temporal constraint:
\[
	z_{i,t} = \sum_{s=1}^{2n_{i,t}^{w}-1}\left(1(s\leq n_{i,t}^{w})\frac{s}{n_{i,t}^{w}}
	+ 1(s>n_{i,t}^{w})\frac{2n_{i,t}^{w}-s}{n_{i,t}^{w}}\right)y_{i,t-s+1}^m + \epsilon_{i,t}^z,
\]
where $\epsilon_{i,t}^z\sim\distn{N}(0, o_{i})$ captures the log-linear approximation error, and we set $o_{i} = 10^{-8}.$

The mixed-frequency VAR is estimated using MCMC methods. In particular, given the model parameters and the common stochastic volatility, we use the proposed sampler as described in Section~\ref{s:methodology} to sample the missing observations of the monthly and quarterly variables. Then, given these missing observations, standard algorithms can be used to sample the model parameters and the common stochastic volatility.\footnote{For example, the common stochastic volatility can be sampled using the methods in \citet{CCM16} or \citet{chan20}. The VAR coefficients is jointly Gaussian and can be sampled jointly or equation by equation as in \citet{CCCM22}.} To gauge the efficiency of the posterior sampler, we compute the inefficiency factors associated with the missing data and the model parameters. All inefficiency factors are less than 100 (see Appendix C for details), and they are comparable to those of large VARs without missing data. 

\begin{figure}[H]
	\centering
	\includegraphics[width=1\textwidth]{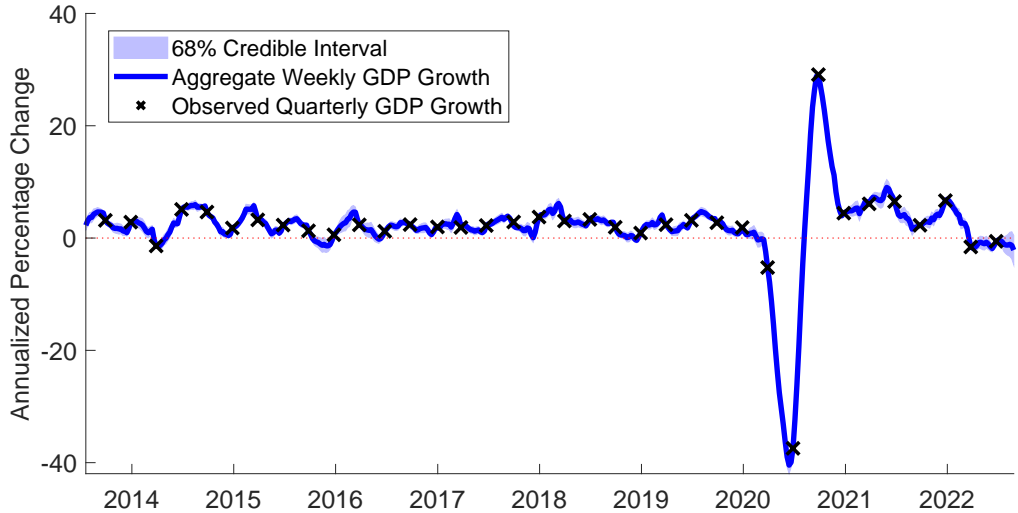}
	\caption{Aggregate weekly estimates of GDP growth over a quarter. The solid line denotes the posterior means of the aggregate weekly GDP growth (annualized), the shaded area is the associated 68\% credible intervals, and the crosses denote the actual observed quarterly GDP values.}
	\label{fig:weeklyGDP}
\end{figure}

The estimated weekly GDP growth rates are rather volatile, which is expected given that they are measured in weekly frequency. For easier comparison and interpretation, we convert these weekly estimates to the more familiar quarterly growth rates. More specifically, given the estimated week-on-week GDP growth rates 
$y_{t,j}^{m}, t=1,\ldots, T,$ we use the inter-temporal constraints to convert them to quarterly growth rates: 
\begin{equation}\label{eq:aggre13}
	y_{t,j}^* = \sum_{s=1}^{n^w_j}\frac{s}{n^w_j}y_{t-s,j}^{m} 
	+ \sum_{s=n^w_j+1}^{2 n^w_j-1}\frac{2n^w_j-s}{n^w_j}y_{t-s,j}^{m},
\end{equation}
where we fix $n^w_j =13$ weeks. Hence, $y_{t,j}^*$ may be interpreted as the cumulative GDP growth over the past 13 weeks. Figure~\ref{fig:weeklyGDP} plots the posterior means and the associated 68\% credible interval of the these aggregate weekly GDP growth rates. In the graph we also mark the observed quarterly GDP growth rates in black crosses. As expected, all the observed quarterly GDP values lie on the aggregate weekly GDP estimates---the inter-temporal constraints ensure that the weekly GDP estimates are aggregated to the observed quarterly value. 

The most prominent feature of the aggregate weekly GDP estimates is the drastic drop at the onset of the COVID-19 pandemic and the subsequent rebound. In particular, the US real GDP decreased by about 37\% in 2020:Q2 when the pandemic forced widespread business closures. When the economy gradually opened up in 2020:Q3, GDP bounced back sharply by about 30\%. One key advantage of modeling GDP in weekly frequency is that, in between the quarterly GDP release dates, the model is able to provide GDP estimates on a weekly basis by incorporating information in other weekly and monthly variables. 

Next, we compare the aggregate weekly GDP estimates to two high-frequency indicators that are designed to track real economic activity. The first is the Weekly Economic Index of \citet{LMST22}, which is updated weekly by the Federal Reserve Bank of New York. The second is the Business Conditions Index of \citet{ADS09}, which is maintained by the Federal Reserve Bank of Philadelphia. While both indicators incorporate a range of macroeconomic and financial variables at high observation frequency, they are latent factors from dynamic factor models. In contrast, our mixed-frequency VAR provides GDP estimates directly and are easier to interpret. 

\begin{figure}[H]
	\centering
	\includegraphics[width=1\textwidth]{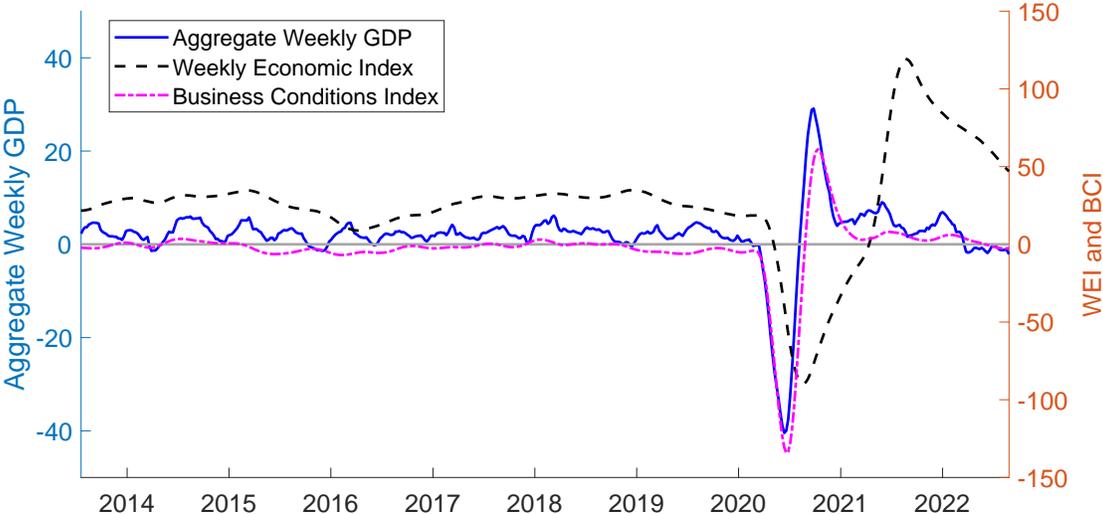}
	\caption{Plots of the aggregate weekly GDP estimates (solid line), the Weekly Economic Index of \citet{LMST22} (dash line) and the Business Conditions Index of \citet{ADS09} (dash-dotted line).}
	\label{fig:weeklyGDP_comparison}
\end{figure}

We obtain the Weekly Economic Index and Business Conditions Index in weekly frequency from the Federal Reserve Banks of New York and Philadelphia, respectively. They are then converted to quarterly growth using~\eqref{eq:aggre13} for easier comparison. Figure~\ref{fig:weeklyGDP_comparison} plots the aggregate weekly GDP estimates as well as the two indicators. It is clear from the figure that the aggregate weekly GDP track the Business Conditions Index closely, even during the extreme turning points in 2020:Q2 and 2020:Q3.\footnote{While the Business Conditions Index is constructed so that its average value is zero, the average GDP growth is about 2\% over the sample period. Hence, there are differences in the level of the two series.} In contrast, the Weekly Economic Index displays noticeably different dynamics during the onset of the COVID-19 pandemic and the immediate rebound. In particular, the Weekly Economic Index suggests that the US economy experienced a sluggish recovery from the widespread lock-down in 2020:Q2. In contrast, the aggregate weekly GDP and the Business Conditions Index indicate a sharper rebound. One potential driver for this difference is that both the aggregate weekly GDP and the Business Conditions Index incorporate quarterly GDP data, whereas the Weekly Economic Index does not. Consequently, the latter could potentially capture only the economic activity of specific sectors of the US economy, whereas the former two track the whole US economy through the information in the GDP data.

%
%
%
%
%
%
\subsection{A Dynamic Factor Model with Unbalanced Datasets}


To demonstrate the versatility of the proposed approach, in the second application we consider a different type of missing data pattern in the context of another popular model for handling large datasets: a dynamic factor model. More specifically, we extract common factors from the FRED-MD datasets of \citet{MN16} using a dynamic factor model with stochastic volatility. We focus on the COVID-19 pandemic period and estimate the common factors in real-time using vintages from March 2020 to September 2022. Each vintage contains 128 monthly variables, but many of these variables have missing values. These missing values mainly come from two sources: missing observations at the beginning of the sample for some more recently constructed variables and missing values at the end of the sample due to publication lags, which is often referred to as ragged edge.

Let $\by_t=(\by_{t}^{o \prime},\by_{t}^{m \prime})'$ denote the $n\times 1$ vector of monthly variables, where $\by_{t}^{o}$ is the $n_t^o$-vector of observed variables and $\by_{t}^{m}$ is the $n_t^m$-vector of variables with missing values at time $t$. Furthermore, let $\bff_t$ represent the $k\times 1$ vector of latent dynamic factors. Due to the COVID-19 outliers, we incorporate stochastic volatility in both the factors and the idiosyncratic errors and consider the following dynamic factor model:
\begin{align*}
	\by_t & = \bA\bff_t + \vepsilon_t, \\	
	\vepsilon_{t} & = \vPsi_1\vepsilon_{t-1} + \cdots + \vPsi_p\vepsilon_{t-p} + \bu_{t}, \quad
		\bu_{t} \sim\distn{N}(\mathbf{0}_n, \vSigma_t),	\\
	\bff_t & = \vPhi_1\bff_{t-1} + \cdots + \vPhi_q\bff_{t-q} + \vepsilon_t^{\bff}, \quad
	\vepsilon_t^{\bff} \sim \distn{N}(\mathbf{0}_k, \vOmega_t),	
\end{align*}
where $\vPsi_1,\ldots, \vPsi_p, \vPhi_1,\ldots, \vPhi_q$ are diagonal matrices, $\vSigma_t = \text{diag}(\e^{h_{1,t}}, \ldots, \e^{h_{n,t}})$ and $\vOmega_t = \text{diag}(\e^{h_{n+1,t}}, \ldots, \e^{h_{n+k,t}})$. The $n+k$ log-volatility processes are assumed to follow independent random walks:
\[
	h_{i,t} = h_{i,t-1} + \epsilon_{i,t}^h, \quad \epsilon_{i,t}^h\sim \distn{N}(0,\sigma^2_{h,i}), \;
	i=1,\ldots, n+k,
\]
where the initial conditions $h_{1,0},\ldots, h_{n+k,0} $ are treated as unknown parameters. Following \citet{ADP21}, we set $p=q=2$.

Using the $PC_p$ criteria proposed in \citet{BN02}, \citet{MN16} find that the optimal number of factors is 8 for their datasets. We therefore set $k=8$. For identification purposes, the factor loading matrix $\bA$ is assumed to be lower triangular with the diagonal elements set to be 1. We then pick the first 8 variables carefully to aid the interpretation of the latent factors. In particular, we draw on the results in \citet{MN16} and use the variables that load most heavily on each of the factors.\footnote{These 8 variables are `usgood', `t10yffm', `cusr0000sac', `aaa', `gs5', `ipcongd', `S\&P: indust' and 'exszusx'. These variables do not have missing values in the vintages we consider.} Their results show that the first factor explains a significant portion of the variation in industrial production and many labor market variables, suggesting that it captures the broad economic conditions. The second factor explains particularly well the variations in interest rate spreads, whereas the third and fourth factors have good explanatory power for variations in prices and interest rates, respectively. 

Next, we specify the prior distributions on the model parameters. Let $\ba$ denote the free elements of the  factor loadings matrix $\bA$, and let $\vpsi$ and $\vphi$ represent the vectors consisting of the diagonal elements of $\vPsi_i,i=1,\ldots, p$ and  $\vPhi_j, j=1,\ldots, q$, respectively. Then, consider the following independent priors on $\ba$, $\vpsi$ and $\vphi$:
\[
	\ba \sim\distn{N}(\ba_{0},\bV_{\ba}), \quad \vpsi\sim\distn{N}(\vpsi_{0},\bV_{\vpsi})1(|\vpsi|<1),
	\quad \vphi\sim\distn{N}(\vphi_{0},\bV_{\vphi})1(|\vphi|<1),	
\]
where the indicator functions ensure the elements $\psi_i$ and $\phi_j$, $i=1,\ldots,np, j=1,\ldots, kq, $ are less then 1 in absolute value. We set the prior means $\ba_{0}, \vpsi_{0}$ and $\vphi_{0}$ to be zero, and the prior covariance matrices to be $\bV_{\ba} = \mathbf{I}_{r}$ with $r=rn-r(r+1)/2$, $\bV_{\vpsi} = 0.01\mathbf{I}_{np}$ and $\bV_{\vphi} = 0.01\mathbf{I}_{nq}$. Finally, for the parameters in the stochastic volatility equations, we assume
 \[
	\sigma_{h,i}^2 \sim \distn{IG}(\nu_{h,i},S_{h,i}), \quad 
	h_{i,0} \sim \distn{N}(m_{h,i},V_{h_{i,0}}), \; i=1,\ldots, n+k,
\]
where we set $m_{h,i} = 0, V_{h_{i,0}} = 0.01$, $\nu_{h,i} = 3$ and $S_{h,i} = 1$ so that the prior means of $h_{i,0}$ and $\sigma^2_{h,i}$ are 0 and $0.5$, respectively. 


This dynamic factor model with an unbalanced panel can be estimated using MCMC methods. More specifically, given the model parameters and the latent factors, we simulate the missing values using the proposed sampler as described in Section~\ref{s:methodology}. Then, given the sampled missing values, the model parameters and latent factors can be drawn from their full conditional distributions using standard algorithms; see, e.g., see \citet{ADP21} and \citet{chan22}. To assess the efficiency of the posterior sampler, we compute the inefficiency factors associated with the missing data and the model parameters (see Appendix C for details). In particular, all inefficiency factors are less than 100, and they are comparable to those of a dynamic factor model with a balanced panel. 

We estimate the dynamic factor model using FRED-MD data vintages from March 2020 to September 2022 that cover the COVID-19 pandemic period. For each data vintage, we first transform the series according to the recommendation in \citet{MN16}. Following common practice, we then standardize each series so that it has 0 mean and unit variance. As mentioned earlier, each vintage contains 128 monthly variables, but many have missing values at the beginning or at the end of the sample. For comparison, we also estimate a version of the dynamic factor model using only variables without missing values, i.e., in each vintage we omit any variables that have missing values. Across the data vintages we consider, on average about 24 variables have missing values and are omitted from the estimation of the factors.

\begin{figure}[H]
	\includegraphics[width=.95\textwidth]{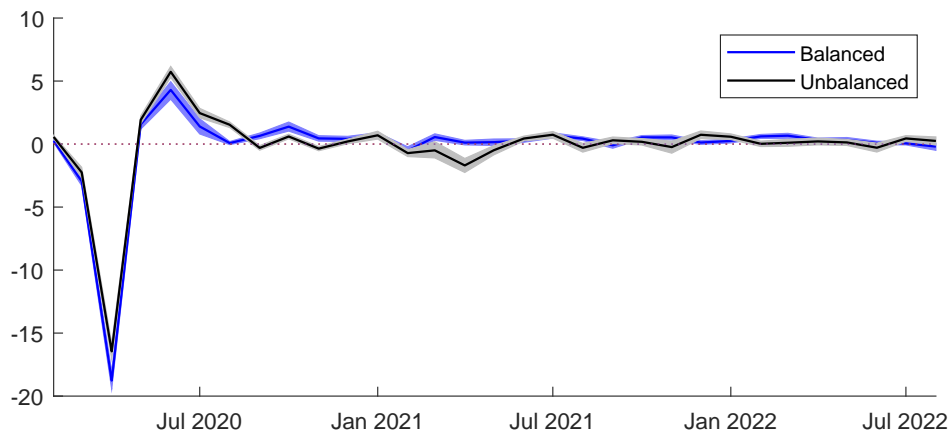}	
	\caption{Filtered estimates of the first factor from the dynamic factor model with balanced and unbalanced datasets. The black line denotes the posterior means of the first factor obtained using all 128 variables (with missing data), and the gray shaded area represents the associated 68\% credible intervals. The blue line
denotes the posterior means obtained using only variables without missing values, and the blue shaded area represents the associated 68\% intervals.}
	\label{fig:factor-1}
\end{figure}

Figure~\ref{fig:factor-1} plots the filtered estimates of the first factor from the dynamic factor model with balanced and unbalanced datasets. These two estimates are broadly similar, and they track well the pronounced downturn at the onset of the COVID-19 pandemic and the subsequent rebound, confirming that the first factor captures the broad economic conditions. However, they also show noticeable differences, especially at the peak and trough associated with the economy-wide reopening after the lock-down. In particular, the factor estimates obtained using all 128 variables show a less severe down-turn ($-16.4$ vs $-18.8$) and a more pronounced uptick afterward ($5.7$ vs $4.3$). These differences may be attributed to the missing values in a number of orders and inventories variables, such as new orders for consumer goods (`acogno') and for capital goods (`andenox') at the beginning of the sample and total business inventories (`businvx') at the end of the sample.

\begin{figure}[H]
	\includegraphics[width=.95\textwidth]{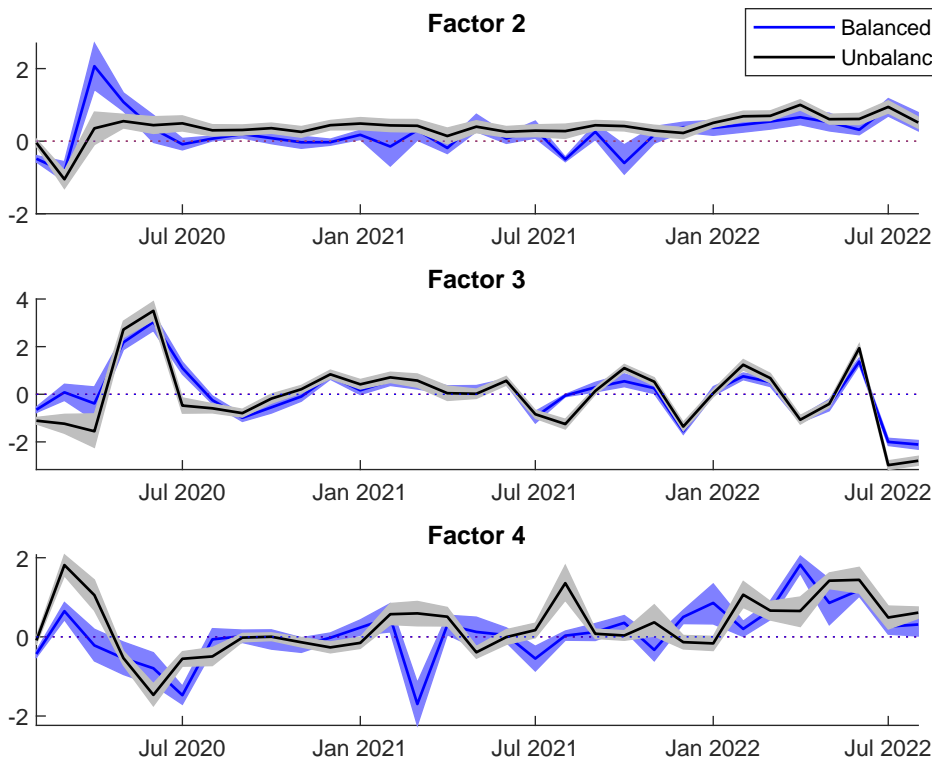}	
	\caption{Filtered estimates of the second, third and fourth factors from the dynamic factor model with balanced and unbalanced datasets. The black line denotes the posterior means of the factor obtained using all 128 variables (with missing data), and the gray shaded area represents the associated 68\% credible intervals. The blue line
denotes the posterior means obtained using only variables without missing values, and the blue shaded area represents the associated 68\% intervals.}
	\label{fig:factor-2-4}
\end{figure}

Next, we report in Figure~\ref{fig:factor-2-4} the filtered estimates of the second, third and fourth factors. There are more substantial differences between the factor estimates obtained using the balanced vs unbalanced datasets. In particular, the most striking differences are those for the fourth factor, which explains variations in interest rates particularly well. These differences might reflect the fact that many of the variables with missing values load heavily on the fourth factor. For example, a number of new private housing permits variables have missing values, and they presumably contain useful information on interest rates. Ignoring those variables is likely to give an incomplete picture on the development of interest rates.\footnote{Filtered estimates of the fifth to eighth factors are reported in Appendix C. We also find substantial differences in the estimates from the balanced vs unbalanced datasets.} 

All in all, these results suggest that omitting variables that have missing values from the empirical analysis can potentially misrepresent the dynamics of broad economic conditions and co-movements in interest rates or prices. This further underlines the utility of the proposed approach to impute missing values that is flexible and works well in large-scale models.

\section{Conclusion} \label{s:conclusion}

We have introduced a novel and efficient approach---that is applicable to any conditionally Gaussian state space models and datasets with arbitrary missing data patterns---for sampling all the missing observations in one step. We have showed via a series of Monte Carlo simulations that the proposed approach is more computationally efficient than standard Kalman-filter based methods under a wide variety of settings. We also demonstrated how the proposed approach can be applied to two empirical macroeconomic applications involving a large mixed-frequency VAR and a dynamic factor model with unbalanced datasets. Both empirical applications illustrated the usefulness of incorporating more information (from high-frequency indicators or variables with missing values) into macroeconomic analysis.


\newpage

\section*{Appendix A: Dataset for the Mixed-Frequency VAR}

This appendix provides details of the 22-variable dataset of the mixed-frequency VAR application. Specifically, Table~\ref{tab:data} describes the 22 variables and their transformations. The first seven weekly variables are obtained from \citet{LMST22}, and the rest are sourced from the FRED database. The sample period is from January 2013 to August 2022.

\begin{table}[H]
\caption{The list of variables and the corresponding transformation used in the mixed-frequency VAR application.}
\label{tab:data}
\resizebox{\textwidth}{!}{\begin{tabular}{lccc}
\hline\hline
Variable 	&	 Frequency 	&	 FRED mnemonic 	&	 Transformation	\\ \hline
Fuels	&	 Weekly 	&	 - 	&	 $100\Delta \text{ln} \left(\frac{x_{t}}{x_{t-52}}\right)$	\\
\rowcolor{lightgray}
Raw Steel Production	&	 Weekly 	&	 - 	&	 $100\Delta \text{ln} \left(\frac{x_{t}}{x_{t-52}}\right)$	\\
Retail Sales Average	&	 Weekly 	&	 - 	&	Level	\\
\rowcolor{lightgray}
Electric Utility Output	&	 Weekly 	&	 - 	&	 $100\Delta \text{ln} \left(\frac{x_{t}}{x_{t-52}}\right)$	\\
US Railroad Traffic	&	 Weekly 	&	 - 	&	 $100\Delta \text{ln} \left(\frac{x_{t}}{x_{t-52}}\right)$	\\
\rowcolor{lightgray}
Continued Claims 	&	 Weekly 	&	 CCSA 	&	 $\text{ln}x_{t}$	\\
Initial Claims 	&	 Weekly 	&	 ICSA 	&	 $\text{ln}x_{t}$	\\
\rowcolor{lightgray}
US Regular All Formulations Gas Price 	&	 Weekly 	&	 GASREGW 	&	 $100\Delta \text{ln} \left(\frac{x_{t}}{x_{t-52}}\right)$	\\
Crude Oil Prices 	&	 Weekly 	&	 WCOILWTICO 	&	 $\text{ln}x_{t}$	\\
\rowcolor{lightgray}
National Financial Conditions Index 	&	 Weekly 	&	 NFCI 	&	 Level	\\
S\&P 500	&	Weekly	&	S\&P 500	&	 $100\Delta \text{ln} \left(\frac{x_{t}}{x_{t-1}}\right)$	\\
\rowcolor{lightgray}
Yield on 1-year U.S. Treasury 	&	 Weekly 	&	 DGS1 	&	 Level	\\
Yield on 10-year U.S. Treasury 	&	 Weekly 	&	 DGS10  	&	 Level	\\
\rowcolor{lightgray}
Moody's Seasoned Baa Corporate Bond Yield 	&	 Weekly 	&	 WBAA  	&	 Level	\\
Moody's Seasoned Aaa Corporate Bond Yield 	&	 Weekly 	&	 WAAA  	&	 Level	\\
\rowcolor{lightgray}
VIX 	&	 Weekly 	&	 VIXCLS  	&	 Level	\\
Industrial Production 	&	 Monthly 	&	 INDPRO  	&	 $100\Delta \text{ln} \left(\frac{x_{t}}{x_{t-1}}\right)$	\\
\rowcolor{lightgray}
Consumer Price Index for All Urban Consumers 	&	 Monthly 	&	 CPIAUCSL  	&	 $100\Delta \text{ln} \left(\frac{x_{t}}{x_{t-1}}\right)$	\\
Unemployment Rate 	&	 Monthly 	&	 UNRATE  	&	Level	\\
\rowcolor{lightgray}
All Employees, Total Nonfarm 	&	 Monthly 	&	 PAYEMS  	&	 $100\Delta \text{ln} \left(\frac{x_{t}}{x_{t-1}}\right)$	\\
Average Weekly Hours: Manufacturing 	&	 Monthly 	&	 AWHMAN  	&	 $\frac{x_{t}}{10}$	\\
\rowcolor{lightgray}
Real Gross Domestic Product 	&	 Quarterly 	&	 GDPC1 	&	 $400\Delta \text{ln} \left(\frac{x_{t}}{x_{t-1}}\right)$	\\ \hline \hline
\end{tabular}
}
\end{table}

\newpage

\section*{Appendix B: The Durbin-Koopman Simulation Smoother}

To implement the Durbin-Koopman simulation smoother, we first need to rewrite the VAR(5) into its companion form:
\begin{equation}\label{eq:DK1}
	\mathbf{s}_{t}=\mathbf{F}_{0}+\mathbf{F}_{1}\mathbf{s}_{t-1} + \bv_{t},\quad 
	\bv_{t} \sim \distn{N}(\mathbf{0}_{5n},\vOmega),
\end{equation}
where $\mathbf{s}_{t} = (\mathbf{y}_{t}^{'},\mathbf{y}_{t-1}^{'},\mathbf{y}_{t-2}^{'},\mathbf{y}_{t-3}^{'},\mathbf{y}_{t-4}^{'})$ is the $5n\times1$ state vector and 
\[
	\vOmega=\left[\begin{array}{cc}
	\vSigma & \mathbf{0}_{n\times 4n}\\
\mathbf{0}_{4n\times n} & \mathbf{0}_{4n\times 4n}
\end{array}\right], \quad
\begin{array}{cc}
\mathbf{F}_{0}=\left[\begin{array}{c}
\mathbf{b}_{0}\\
\mathbf{0}_{4n\times1}
\end{array}\right], & \mathbf{F}_{1}=\left[\begin{array}{ccccc}
\mathbf{B}_{1} & \mathbf{B}_{2} & \mathbf{B}_{3} & \mathbf{B}_{4} & \mathbf{B}_{5} \\
\mathbf{I}_{n} & \mathbf{0}_{n} & \mathbf{0}_{n} & \mathbf{0}_{n} & \mathbf{0}_{n} \\
\mathbf{0}_{n} & \mathbf{I}_{n} & \mathbf{0}_{n} & \mathbf{0}_{n} & \mathbf{0}_{n} \\
\mathbf{0}_{n} & \mathbf{0}_{n} & \mathbf{I}_{n} & \mathbf{0}_{n} & \mathbf{0}_{n} \\
\mathbf{0}_{n} & \mathbf{0}_{n} & \mathbf{0}_{n} & \mathbf{I}_{n} & \mathbf{0}_{n} \\
\end{array}\right].\end{array}
\]
When both the high- and low-frequency variables are observed, the measurement equation is given by
\begin{equation}
\mathbf{y}_{t} = \vLambda_{1}\mathbf{s}_{t}. \label{eq:DK2}
\end{equation}
When only the high-frequency variables are observed, it becomes
\begin{equation}
	\mathbf{y}_{t}^{o} = \vLambda_{2}\mathbf{s}_{t}.\label{eq:DK3}
\end{equation}
The matrices $\vLambda_{1}$ and $\vLambda_{2}$ incorporate the inter-temporal constraints:
\begin{align*}
	\vLambda_{1} & =\left[\begin{array}{cccccccccc}
\mathbf{I}_{n^{o}} & \mathbf{0}_{n^{m}} & \mathbf{0}_{n^{o}} & \mathbf{0}_{n^{m}} & \mathbf{0}_{n^{o}} & \mathbf{0}_{n^{m}} & \mathbf{0}_{n^{o}} & \mathbf{0}_{n^{m}} & \mathbf{0}_{n^{o}} & \mathbf{0}_{n^{m}}\\
\mathbf{0}_{n^{m}} & \frac{1}{3}\mathbf{I}_{n^{m}} & \mathbf{0}_{n^{o}} & \frac{2}{3}\mathbf{I}_{n^{m}} & \mathbf{0}_{n^{o}} & \mathbf{I}_{n^{m}} & \mathbf{0}_{n^{o}} & \frac{2}{3}\mathbf{I}_{n^{m}} & \mathbf{0}_{n^{o}} & \frac{1}{3}\mathbf{I}_{n^{m}}
\end{array}\right], \\
\vLambda_{2} & =\left[\begin{array}{cccccccccc}
\mathbf{I}_{n^{o}} & \mathbf{0}_{n^{m}} & \mathbf{0}_{n^{o}} & \mathbf{0}_{n^{m}} & \mathbf{0}_{n^{o}} & \mathbf{0}_{n^{m}} & \mathbf{0}_{n^{o}} & \mathbf{0}_{n^{m}} & \mathbf{0}_{n^{o}} & \mathbf{0}_{n^{m}}\end{array}\right].
\end{align*}

We follow \citet{Jarocinski15} and implement the Durbin-Koopman simulation smoother on (\ref{eq:DK1}), (\ref{eq:DK2}) and (\ref{eq:DK3}). 

\newpage

\section*{Appendix C: Additional Results}

\begin{figure}[H]
	\includegraphics[width=.95\textwidth]{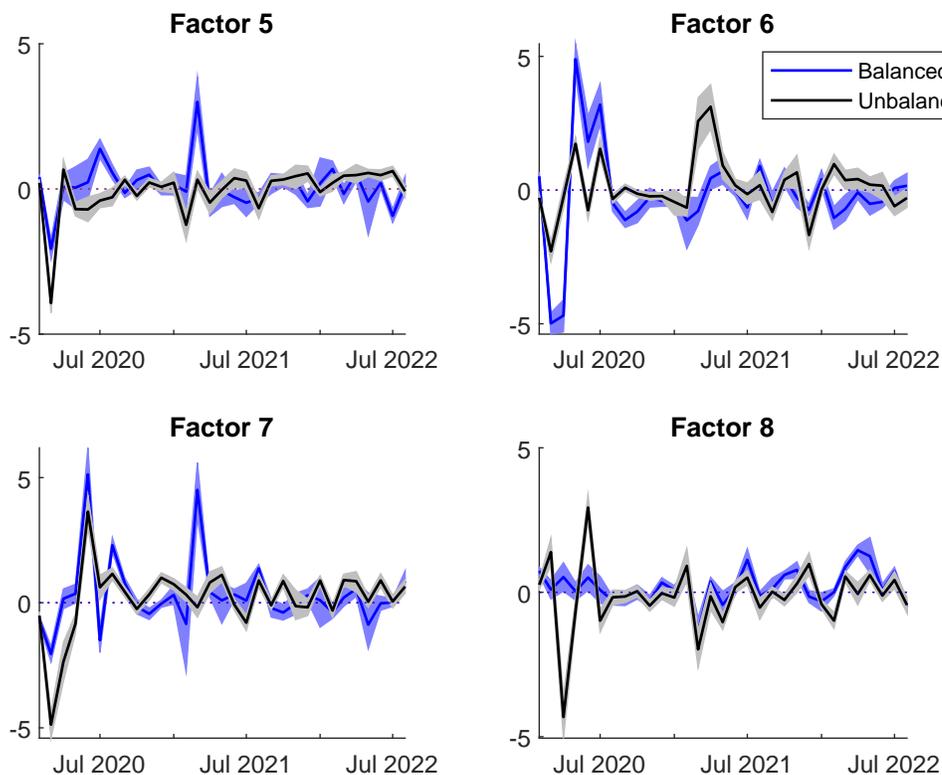}	
	\caption{Filtered estimates of the fifth to eighth factors from the dynamic factor model with balanced and unbalanced datasets. The black line denotes the posterior means of the factor obtained using all 128 variables (with missing data), and the gray shaded area represents the associated 68\% credible intervals. The blue line
denotes the posterior means obtained using only variables without missing values, and the blue shaded area represents the associated 68\% intervals.}
	\label{fig:factor-5-8}
\end{figure}

This appendix presents additional empirical results from the two applications. We first report in Figure~\ref{fig:factor-5-8} the filtered estimates of factors 5-8 in the dynamic factor model using both balanced and unbalanced FRED-MD datasets.

\begin{figure}[H]
	\centering
	\includegraphics[width=.65\textwidth]{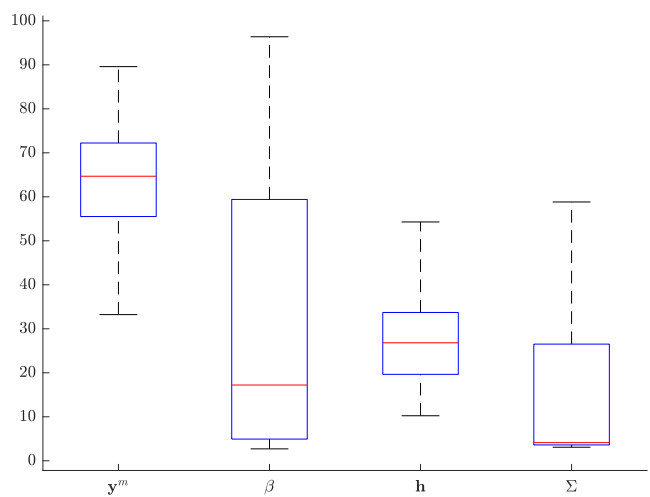}	
	\caption{Boxplots of the inefficiency factors corresponding to the posterior draws of the missing data $\by^{m}$ and other model parameters, $\vbeta, \bh$ and $\vSigma$, from the weekly state-space mixed-frequency VAR.}
	\label{fig:weeklyGDP_IE}
\end{figure}

\begin{figure}[H]
	\centering
	\includegraphics[width=.65\textwidth]{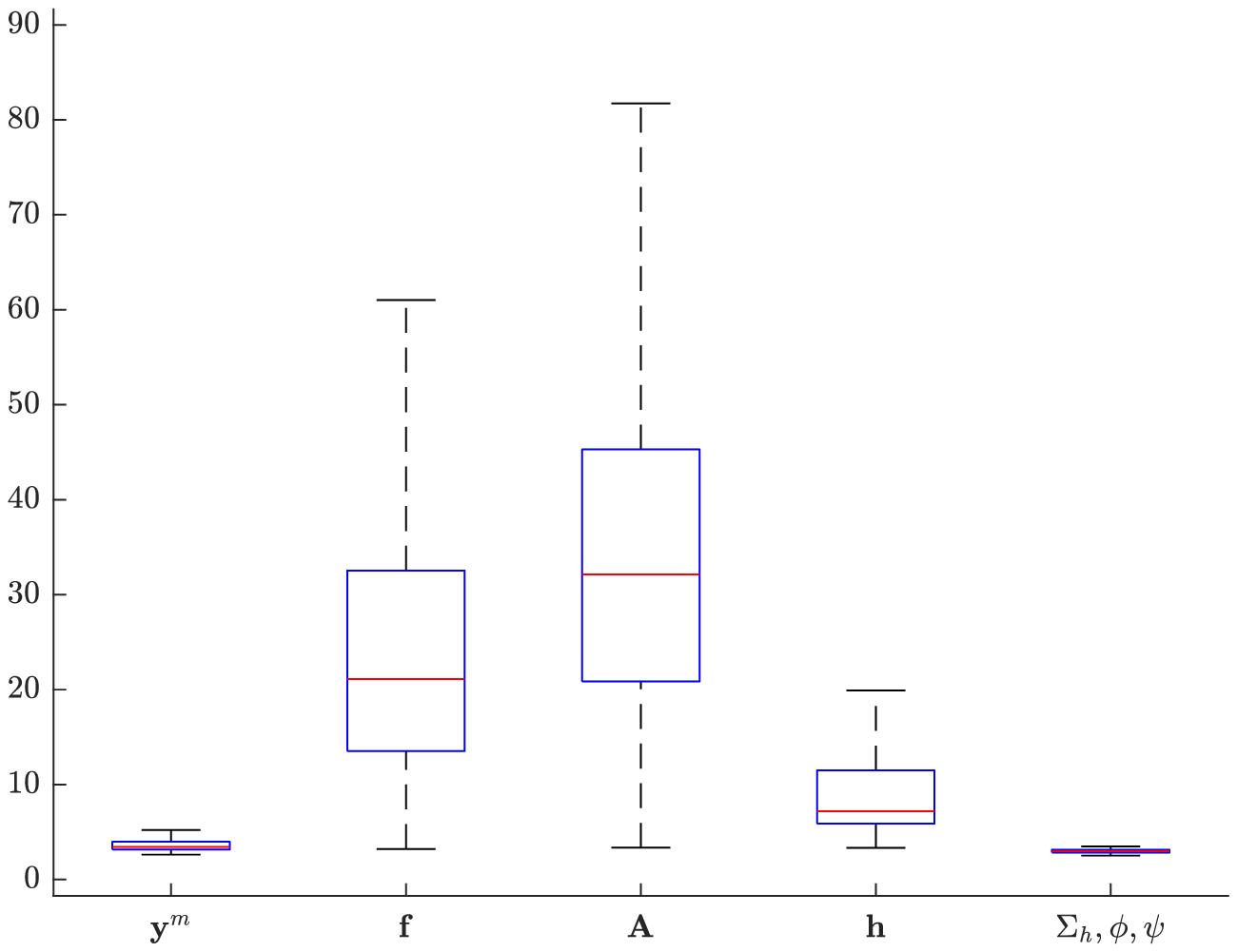}		
	\caption{Boxplots of the inefficiency factors corresponding to the posterior draws of the missing data $\by^{m}$ and other model parameters, $\bff$, $\bA$, $\bh$, $\vsigma^2_{h}, \vpsi$ and $\vphi$ from the dynamic factor model with an unbalanced panel.}
	\label{fig:DFM_unbalanced_IE}
\end{figure}

\begin{figure}[H]
	\centering
	\includegraphics[width=.65\textwidth]{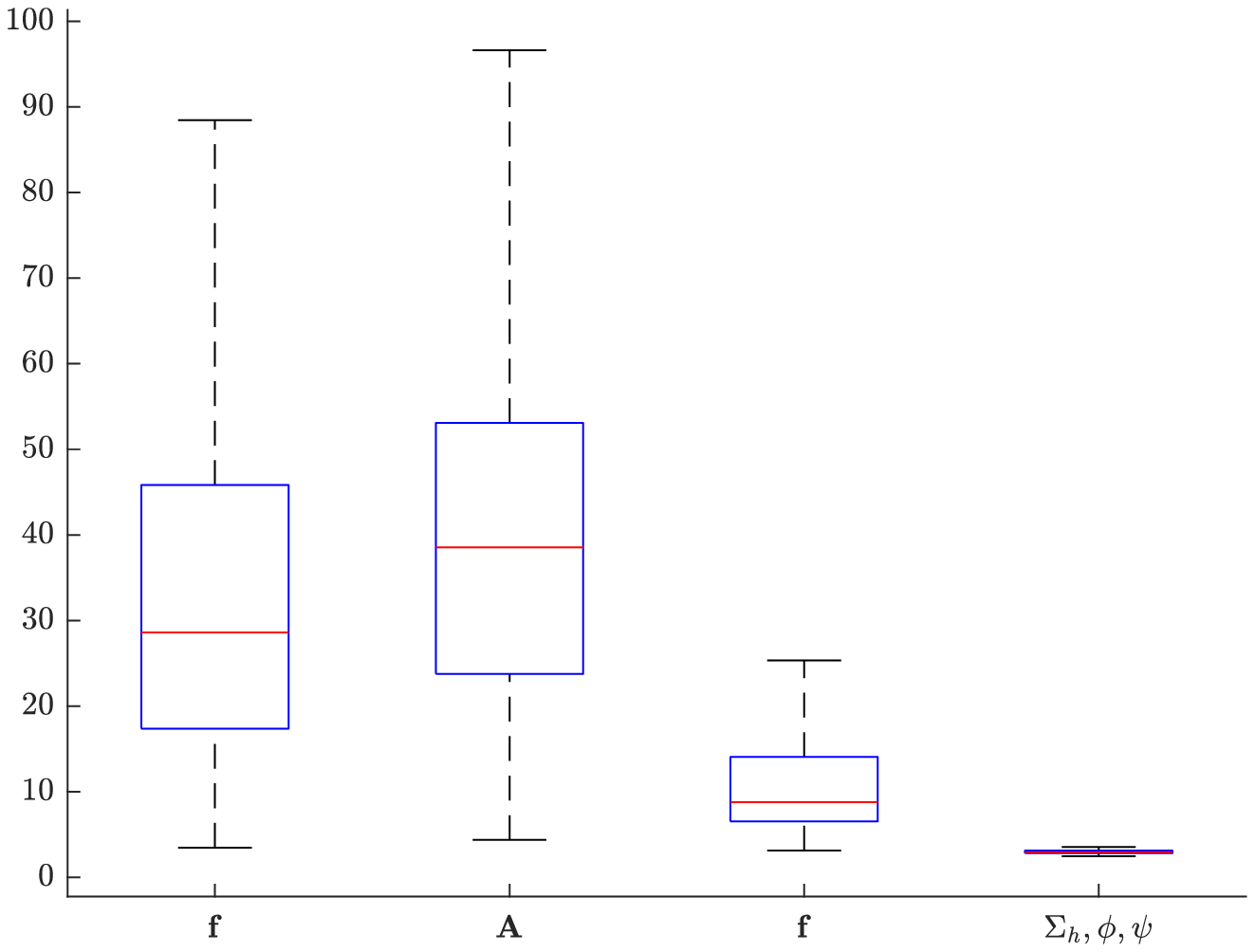}	
	\caption{Boxplots of the inefficiency factors corresponding to the posterior draws of $\bff$, $\bA$, $\bh$, 
	$\vsigma^2_{h}$, $\vpsi$ and $\vphi$ from the dynamic factor model with a balanced panel.}
	\label{fig:DFM_balanced_IE}
\end{figure}

Next, Figures~\ref{fig:weeklyGDP_IE}-\ref{fig:DFM_balanced_IE} report the inefficiency factors of the MCMC samples from the two empirical applications. Specifically, Figure~\ref{fig:weeklyGDP_IE} plots the inefficiency factors of the missing data and other model parameters from the weekly state-space mixed-frequency VAR. To present the information more succinctly, boxplots of the inefficiency factors are used. The middle line of each box denotes the median, while the
lower and upper lines represent, respectively, the 25- and the 75-percentiles. The whiskers
extend to the maximum and minimum. 

Figures~\ref{fig:DFM_unbalanced_IE} and \ref{fig:DFM_balanced_IE} plot the inefficiency factors from the dynamic factor model with an unbalanced panel and a balanced panel, respectively.

\newpage

\singlespace

\ifx\undefined\BySame
\newcommand{\BySame}{\leavevmode\rule[.5ex]{3em}{.5pt}\ }
\fi
\ifx\undefined\textsc
\newcommand{\textsc}[1]{{\sc #1}}
\newcommand{\emph}[1]{{\em #1\/}}
\let\tmpsmall\small
\renewcommand{\small}{\tmpsmall\sc}
\fi

\end{document}